\documentclass[a4paper,fleqn,usenatbib]{mnras}

\usepackage{graphicx}	% Including figure files
\usepackage{amsmath,amsfonts,amssymb}
\usepackage{makeidx}
\usepackage{epstopdf} 
\usepackage{booktabs,caption,fixltx2e}
\usepackage[flushleft]{threeparttable}
\usepackage{hyperref}
\usepackage{multirow}
\usepackage{rotating}
\usepackage{color}
\usepackage[T1]{fontenc}
\usepackage{ae,aecompl}
\usepackage{soul}

\def \BE{\begin{equation}}
\def \EE{\end{equation}}	
\def \BC{\begin{center}}
\def \EC{\end{center}}
\def \BEA{\begin{eqnarray}}
\def \EEA{\end{eqnarray}}

\title[High Frequency Cluster Radio Galaxies]{High Frequency Cluster Radio Galaxies: Luminosity Functions and Implications for SZE Selected Cluster Samples}

\newcommand{\Munich}{$^{1}$}
\newcommand{\ExcellenceCluster}{$^{2}$}
\newcommand{\MPE}{$^{3}$}
\newcommand{\AAUChicago}{$^{4}$}
\newcommand{\KICPChicago}{$^{5}$}
\newcommand{\FNAL}{$^{6}$}
\newcommand{\ANL}{$^{7}$}
\newcommand{\Berkeley}{$^{8}$}
\newcommand{\McGill}{$^{9}$}
\newcommand{\MIT}{$^{10}$}
\newcommand{\UM}{$^{11}$}
\newcommand{\EFIChicago}{$^{12}$}
\newcommand{\PhysicsUChicago}{$^{13}$}

\author[Gupta et al.] {N.~Gupta\thanks{Nikhel.Gupta@Physik.LMU.de}\Munich$^,$\ExcellenceCluster$^,$\MPE,
A.~Saro\Munich$^,$\ExcellenceCluster,
J.~J.~Mohr\Munich$^,$\ExcellenceCluster$^,$\MPE,
B.~A.~Benson\AAUChicago$^,$\KICPChicago$^,$\FNAL,
S.~Bocquet\KICPChicago$^,$\ANL$^,$\Munich$^,$\ExcellenceCluster,
\newauthor 
R.~Capasso\Munich$^,$\ExcellenceCluster,
J.~E.~Carlstrom\AAUChicago$^,$\KICPChicago$^,$\ANL$^,$\EFIChicago$^,$\PhysicsUChicago,
I.~Chiu\Munich$^,$\ExcellenceCluster,
T.~M.~Crawford\AAUChicago$^,$\KICPChicago,
\newauthor
T.~de~Haan\Berkeley$^,$\McGill,
J.~P.~Dietrich\Munich$^,$\ExcellenceCluster,
C.~Gangkofner\Munich$^,$\ExcellenceCluster,
W.~L.~Holzapfel\Berkeley,
\newauthor
M.~McDonald\MIT,
D.~Rapetti\Munich$^,$\ExcellenceCluster,
C.~L.~Reichardt\UM
\\
\\
\Munich Faculty of Physics, Ludwig-Maximilians-Universit\"{a}t, Scheinerstr.\ 1, 81679 Munich, Germany \\
\ExcellenceCluster Excellence Cluster Universe, Boltzmannstr.\ 2, 85748 Garching, Germany \\
\MPE Max Planck Institute for Extraterrestrial Physics, Giessenbachstr.\ 85748 Garching, Germany \\
\AAUChicago Department of Astronomy and Astrophysics, University of Chicago, 5640 South Ellis Avenue, Chicago, IL 60637 \\
\KICPChicago Kavli Institute for Cosmological Physics, University of Chicago, 5640 South Ellis Avenue, Chicago, IL 60637 \\
\FNAL Center for Particle Astrophysics, Fermi National Accelerator Laboratory, Batavia, IL, USA 60510 \\
\ANL Argonne National Laboratory, 9700 S. Cass Avenue, Argonne, IL, USA 60439 \\
\Berkeley Department of Physics, University of California, Berkeley, CA 94720 \\
\McGill Department of Physics,McGill University, 3600 Rue University, Montreal, Quebec H3A 2T8, Canada \\
\MIT Kavli Institute for Astrophysics and Space Research, Massachusetts Institute of Technology, 77 Massachusetts Avenue, \\
Cambridge, MA 02139 \\
\UM School of Physics, University of Melbourne, Parkville, VIC 3010, Australia \\
\EFIChicago Enrico Fermi Institute, University of Chicago, 5640 South Ellis Avenue, Chicago, IL 60637 \\
\PhysicsUChicago Department of Physics, University of Chicago, 5640 South Ellis Avenue, Chicago, IL 60637 \\
}

\begin{document}
\date{Accepted ???. Received ???; in original form ???} 

\maketitle

\begin{abstract}
We study the overdensity of point sources in the direction of X-ray-selected galaxy clusters from the Meta-Catalog of X-ray detected Clusters of galaxies (MCXC; $\langle z \rangle = 0.14$) at South Pole Telescope (SPT) and Sydney University Molonglo Sky Survey (SUMSS) frequencies.  
%We study the 95~GHz and 150~GHz luminosity functions (LFs) of cluster radio galaxies and their surface density profiles within galaxy clusters. We construct LFs from the overdensity of point sources in South Pole Telescope (SPT) data in the direction of X-ray-selected galaxy clusters from the Meta-Catalog of X-ray detected Clusters of galaxies (MCXC; $< z > = 0.14$). 
Flux densities at 95, 150 and 220~GHz are extracted from the 2500~deg$^2$ SPT-SZ survey maps at the locations of SUMSS sources, producing a multi-frequency catalog of radio galaxies.
% detected in the Sydney University Molonglo Sky Survey (SUMSS). 
In the direction of massive galaxy clusters, the radio galaxy flux densities at 95 and 150~GHz are biased low by the cluster Sunyaev-Zel'dovich Effect (SZE) signal, which is negative at these frequencies. We employ a cluster SZE model to remove the expected flux bias and then study these corrected source catalogs.  We find that the high frequency radio galaxies are centrally concentrated within the clusters and that their luminosity functions (LFs) exhibit amplitudes that are characteristically an order of magnitude lower than the cluster LF at 843~MHz.  We use the 150~GHz LF to estimate the impact of cluster radio galaxies on an SPT-SZ like survey.  The radio galaxy flux typically produces a small bias on the SZE signal and has negligible impact on the observed scatter in the SZE mass-observable relation.  If we assume there is no redshift evolution in the radio galaxy LF then $1.8\pm0.7$~percent of the clusters with detection significance $\xi \geq 4.5$ would be lost from the sample.  Allowing for redshift evolution of the form $(1+z)^{2.5}$ increases the incompleteness to $5.6\pm1.0$~percent.  Improved constraints on the evolution of the cluster radio galaxy LF require a larger cluster sample extending to higher redshift.
\end{abstract}

%(NFW concentration $c=108^{+107}_{-48}$), 


\section{Introduction}

The first galaxy cluster sample selected through the Sunyaev-Zel'dovich Effect \citep[SZE;][]{sunyaev72} emerged in the last decade \citep{staniszewski09}; since then, high frequency mm-wave surveys by the South Pole Telescope \citep[SPT;][]{carlstrom11}, the Atacama Cosmology Telescope \citep[ACT;][]{fowler07}, and Planck \citep{planck11-13} have enabled the SZE selection of large cluster samples and their use to constrain cosmological parameters \citep{vanderlinde10, sehgal11, benson13, reichardt13, hasselfield13, bocquet15, planck15, dehaan16}.  In these analyses the connection between the cluster SZE signature and the underlying halo mass -- the so-called mass-observable relation -- plays a central role.  Emission from cluster radio galaxies will contaminate the cluster SZE signature at some level, resulting in incompleteness in the SZE selected cluster samples and contributing to the scatter in the SZE mass-observable relation.  Although previous studies indicate that these effects are small at high frequencies \citep{lin07,lin09,sehgal10,linhenry15}, these studies all rely to some degree on extrapolations from the properties of radio galaxies at low frequencies.
%where SZE flux is insignificant. 

%If not accounted for, these effects could in principle impact the accuracy and precision of cosmological analyses of these samples.  \cite{linhenry15} performed Monte Carlo simulations and mock observations of galaxy clusters and found that none of the observed massive elliptical galaxies \citep{dunn10} and radio-loud AGN \citep{sun09b, hlavacek12} would produce an SZE bias in excess of 10\% if located in massive clusters. However, their measurements are strongly dependent on the choice of spectral index, due to unavailability of any published study of radio galaxies in clusters extending to 150~GHz. 

One way to study this phenomenon more directly is to statistically examine the radio galaxy population using a cluster sample selected in a manner that would be unaffected by galaxy radio emission.  In this work we carry out the first such study at high frequencies,  constructing the cluster radio galaxy luminosity function (LF)  from the overdensity of point sources in the SPT-SZ survey \citep{carlstrom11, bleem15} and the Sydney University Molonglo Sky Survey \citep[SUMSS][]{bock99, mauch03, murphy07} toward galaxy clusters in the Meta-Catalog of X-ray detected Clusters of galaxies \citep[MCXC,][]{piffaretti11}. 

The measurement of the LF is not straightforward at the 95 and 150~GHz SPT observing frequencies, because of the presence of the negative SZE signature at these frequencies. The cluster SZE signature biases our radio galaxy flux measurements, and could indeed remove point sources from a high frequency selected sample.  Thus, to estimate the true underlying radio galaxy flux, one must estimate the cluster SZE flux at the positions of these point sources and then use that to correct the radio galaxy fluxes.  These corrections will lead to additional point sources in a flux limited sample and are therefore crucial for the LF analysis.  We use the SUMSS radio galaxies, observed at 843~MHz, to enable this correction and the construction of unbiased radio galaxy samples.  Specifically, we measure the SPT point source fluxes at the locations of all SUMSS radio galaxies and apply an estimated SZE correction.  Using this corrected catalog we then measure the LFs and use that information to estimate the impact of the cluster radio galaxies on cluster samples selected using high frequency observations of the SZE.  

The plan of the paper is as follows: In section~\ref{sec:data}, we discuss the observations and the data used in this work and describe the corrections applied to the point source catalogs at 95 and 150~GHz. Section~\ref{sec:results} is dedicated to the studies of surface density profiles and the LFs. In section~\ref{sec:contamination} we estimate the contamination by radio galaxies in SZE cluster surveys.  Section~\ref{sec:systematics} describes the effect of cluster mass and point source flux uncertainties on our results.  We conclude in section~\ref{sec:conclusions}. Throughout this paper we assume a flat $\Lambda$CDM cosmology with matter density parameter $\Omega_{\rm M} = 0.3$ and Hubble constant $H_0$ = 70~$\rm km$ $\rm s^{-1}$ $\rm Mpc^{-1}$.  We take the normalization of the matter power spectrum to be $\sigma_8= 0.83$.
\section{Data and Radio Galaxy Flux Corrections}
\label{sec:data}

We study the overdensity of radio point sources in the direction of galaxy clusters in the MCXC. The radio sources are selected from the SUMSS catalog observed at 843~MHz and  SPT observations are used to measure the source fluxes at higher frequencies. We discuss these observations in the following sections. At 95 and 150~GHz frequencies, the source fluxes in clusters can be biased by their SZE flux. Taking this into account we construct an unbiased catalog of SUMSS sources at high frequencies using the independently detected SPT point source catalog as well as SPT-SZ maps, as described in section~\ref{sec:corrections}.
\subsection{SPT Observations}
\label{sec:SPT}

The South Pole Telescope (SPT) is a 10-meter telescope located at the Amundsen-Scott South Pole station in Antarctica \citep{carlstrom11}.  The 2500~deg$^2$ SPT-SZ survey has coverage in multiple frequency bands centered around 95, 150 and 220~GHz, corresponding to wavelengths of 3.2, 2.0 and 1.4 mm, respectively.  The SPT angular resolution at these three frequencies is approximately $1\farcm6$, $1\farcm1$ and $1\farcm0$, and the survey depths are approximately 40, 18 and 70 $\mu$K-arcmin, respectively.
 
The data reduction procedure for SPT is described in detail elsewhere \citep{staniszewski09,vieira10,schaffer11,mocanu13}.  To increase the signal-to-noise ratio ($S/N$) of unresolved objects a matched filter $\psi$ \citep{tegmark98} of the following form is generated
\BE
\label{eqn:matched_filter}
\psi \equiv \frac{\tau^T N^{-1}}{\sqrt{\tau^T N^{-1} \ \tau}},
\EE
where $\tau$ is the source shape, which is a function of the beam and filtering, and $N$ is the noise covariance matrix, which also includes astrophysical contaminants like primary CMB anisotropy along with the instrument and atmospheric noise. The purpose of this filtering is to increase the sensitivity of the beam size objects by down-weighting signal from larger and smaller scales where the $S/N$ is small. 

Sources in the filtered SPT-SZ maps were identified using the CLEAN algorithm \citep{hogbom74}. We again refer the reader to \citet{vieira10} and \citet{mocanu13} for details about the implementation of the CLEAN algorithm to the SPT maps. The flux of the identified sources is calculated from the filtered maps by converting the value of the brightest pixel across the sources from the units of CMB fluctuation temperature to the flux density as follows
\BE
\label{eqn:temp_to_flux}
 S[\mathrm{Jy}] = T_\mathrm{peak} \cdot \Delta \Omega_\mathrm{f} \cdot 10^{26} \cdot \frac{2k_{\rm B}}{c^2} \left ( \frac{k_{\rm B} T_{\rm CMB}}{h} \right )^2 \frac{x^4 e^x}{(e^x-1)^2},
\EE  
where $x=h\nu/(k_{\rm B} T_{\rm CMB})$, $T_\mathrm{peak}$ is the peak temperature in a pixel, $k_{\rm B}$ is the Boltzmann constant, $c$ is the velocity of light, $T_{\rm CMB}$ is the present CMB temperature, $h$ is the Planck constant and $\Delta \Omega_\mathrm{f}$ is the effective solid angle under the source template. There are 4841 point sources detected by SPT above a $S/N$ of 4.5 in any of the three frequency bands \citep{everett16}.  Of these, we expect $\sim$80 percent to have synchrotron dominated emission and the rest to be dusty galaxies, consistent with the findings in the analysis of 771~deg$^2$ of the SPT-SZ survey \citep{mocanu13}.

\subsection{SUMSS Catalog}
\label{sec:SUMSS}
The Sydney University Molonglo Sky Survey \citep[SUMSS,][]{bock99, mauch03, murphy07} imaged the southern radio sky at 843~MHz using the Molonglo Observatory Synthesis Telescope \citep[MOST,][]{mills81, robertson91}. The survey was completed in early 2007 and covers 8100 $\rm deg^2$ of sky with $\delta$$\leq$$-30^\circ$ and $\vert b \vert$$\geq$$10^\circ$. The catalog contains 210,412 radio sources to a limiting peak brightness of 6 mJy beam$^{-1}$ at $\delta$$\leq$$-50^\circ$ and 10 mJy beam$^{-1}$ at $\delta$$>$$-50^\circ$.  At the SUMSS selection frequency, we expect nearly all sources above the flux selection threshold to be synchrotron dominated \citep{dezotti05}. The position uncertainties in the catalog are always better than 10$^{\prime\prime}$. In fact, for sources with peak brightness $A_{843}$$\geq$20 mJy beam$^{-1}$, the accuracy is in the range 1$^{\prime\prime}$ to 2$^{\prime\prime}$. The flux density measurements are accurate to within 3~percent. The catalog is complete to 8 mJy at $\delta$$\leq$$-50^\circ$ and to 18 mJy at $\delta$$>$$-50^\circ$. There are approximately 56,000 SUMSS sources in the SPT region at 100 percent completeness.
%55884 is the actual number Nikhel had here

\subsection{MCXC Catalog}
\label{sec:MCXC}
For our analysis, we use the Meta-Catalog of X-ray detected Clusters of galaxies \citep[MCXC,][]{piffaretti11}, which is compiled from the publicly available ROSAT All Sky Survey-based catalogs, such as, NORAS \citep{bohringer00}, REFLEX \citep{bohringer04}, BCS \citep{ebeling98, ebeling00}, SGP \citep{cruddace02}, NEP \citep{henry06}, MACS \citep{ebeling01}, CIZA \citep{ebeling02, kocevski07} and serendipitous catalogs such as, 160SD \citep{mullis03}, 400SD \citep{burenin07}, SHARC \citep{romer00}, WARPS \citep{perlman02, horner08a}, and EMSS \citep{gioia94, henry04}. The catalog contains a total of 1,743 clusters in the whole sky. The cluster coordinates are those of the cluster centroid determined from X-ray data (apart from the 47 clusters in the sub-catalogue EMSS \citep{gioia94} which have the coordinates of the cluster optical position). The masses are estimated from the homogenized luminosities using the power law relation described in \citet{piffaretti11}. The redshift range of the MCXC catalog spans from 0.003 to 1.26 with a median of 0.14, and the mass range is $9.6\times10^{11}\rm{M_{\odot}}$$\le$$M_{500}$$\le$$2.2\times10^{15}\rm{M_{\odot}}$ with a median mass of $1.76\times10^{14}\rm{M_{\odot}}$. 
%Unless stated, we always use the critical density as the measure of cluster mass in this paper. 
Here $M_{500}$ describes the mass of the cluster within the sphere where the density is 500 times the critical density of the Universe. We use an NFW profile with the expected concentration from large scale structure simulations to convert from $M_{500}$ to $M_{200}$ \citep{duffy08,navarro97}. 

There are 139 and 333 MCXC clusters in the SPT-SZ and SUMSS regions, respectively.  In the SPT-SZ region these systems span a mass range $6.5\times10^{12}\rm{M_{\odot}}$$\le$$M_{500}$$\le$$1.2\times10^{15}\rm{M_{\odot}}$ with a median mass of $1.5\times10^{14}\rm{M_{\odot}}$.  In the SUMSS region the corresponding mass range and median mass are $6.5\times10^{12}\rm{M_{\odot}}$$\le$$M_{500}$$\le$$1.2\times10^{15}\rm{M_{\odot}}$ and $1.8\times10^{14}\rm{M_{\odot}}$, respectively.  The median redshift for both samples is $z\sim0.1$, and the highest redshift system is at $z=0.686$.  So, using the MCXC cluster sample to identify cluster radio galaxies allows us to examine primary low redshift systems that cover the mass range from groups to clusters.

\begin{figure}
\centering
\includegraphics[width=9cm, height=7.5cm]{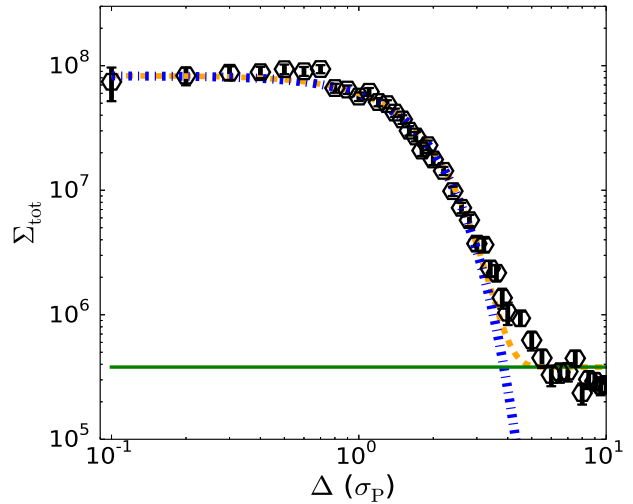}
\vskip-0.1in
\caption{Distribution of offsets between SUMSS and SPT point sources in units of the total positional uncertainty $\sigma_{\rm P}$.  Lines represent best fit level of random associations (green), Gaussian core (blue) and core plus random (orange).  We limit matches to lie within 5$\sigma_{\rm P}$ and estimate only 3~percent of those are random associations.}
\label{fig:SUMSS_SPT}
\end{figure}

\subsection{Catalog of Cluster Radio Galaxy Candidates}
\label{sec:corrections}

The intrinsic flux of a point source residing along the line of sight to a galaxy cluster is biased by the cluster SZE flux.  The SZE flux corrections are expected to be small for non-central sources, but when one defines a radio galaxy sample using a flux limit the presence of this SZE flux bias inevitably means that some sources that should be in the flux limited catalog will drop out of it.  Thus, we employ the SUMSS catalog in building a catalog of high frequency radio galaxies.

As described in the following sections, this requires matching the SUMSS and SPT catalogs for the subset of radio galaxies that are bright enough to have made it into the SPT catalog and extracting a flux density measurement directly from the appropriately filtered SPT maps for the rest of the sources.  We describe here the results of the catalog matching, the SZE flux bias correction and then the characteristics of the final analysis-ready SUMSS selected catalog at SPT frequencies.

\subsubsection{Matching SUMSS and SPT Sources}
\label{sec:sumss_plus_spt}

The $S/N$ and fluxes of 4,841 SPT detected point sources are measured at 95, 150 and 220~GHz with the methodology described in section~\ref{sec:SPT}. The positional uncertainty $\sigma_\mathrm{SPT}$ along one dimension of these point sources depends upon the $S/N$ of the sources as well as the beam size $\sigma_{\rm F}$ \citep{ivison07}, where FWHM$=\sqrt{8 \ln 2}~\sigma_\mathrm{F}$, and is given by
%
%  FWHM is the resilution 1.1' at 150 GHz.
%
\BE
\label{eqn:pos_uncertainty0}
\sigma_{\mathrm {SPT}} = \sqrt{2}  \frac{\sigma_{\mathrm F}}{S/N}.
%\sigma_{\mathrm {SPT}} = \frac{\sqrt{2} \sigma_{\mathrm F}}{S/N}.
\EE
We choose the smallest value of $\sigma_{\rm {SPT}}$ out of the three SPT bands as the positional uncertainty of each point source. 

To select the radio galaxy population, we look for the SUMSS counterparts within a region of radius 5$\sigma_{\rm p}$ around the SPT selected point sources, where $\sigma_{\rm p}$ is the quadrature sum of the SPT and SUMSS positional uncertainties
\BE
\label{eqn:pos_uncertainty}
\sigma_{\mathrm p}^2 = \sigma_{\mathrm {SPT}}^2 + \sigma_{\rm SUMSS}^2.
\EE
We choose the 5$\sigma_{\rm p}$ limit to search for SUMSS counterparts, because the surface density of the SUMSS sources within distance $\Delta\sigma_{\rm p}$ of SPT point sources drop to a uniform background level at $\Delta\sigma_{\rm p}\sim5$ (see Fig.~\ref{fig:SUMSS_SPT}). We find 3,558 (72~percent) of the SPT detected point sources to have SUMSS counterparts. This fraction is similar to the 71~percent of sources with SUMSS counterparts found by \citet{mocanu13} in the first 771~$\deg^2$ of the SPT-SZ region.

We fit a Gaussian model along with a constant background to the surface density profile of SUMSS sources in SPT as shown in Fig.~\ref{fig:SUMSS_SPT}.  Using this fit we estimate that the purity of the sample selected within 5$\sigma_{\rm p}$ is 97~percent. 

\subsubsection{SZE Flux Bias and Correction}
\label{sec:SZEcorrections}

The flux of sources in the direction of galaxy clusters is suppressed by the negative SZE signature. Thus, to recover the true flux we need to correct for the SZE flux bias.  To do this, we first create an SZE map of the overlapping cluster using a circularly symmetric Compton $Y$ profile \citep{arnaud10} extending to a radius $5R_{200}$, appropriate for a cluster of the mass given in the MCXC catalog.  Specifically, SZE maps with the same pixel size as the SPT maps are created by scaling the $Y$ signal in a pixel by the pixel area.  We then filter these cluster maps using the matched filter technique for unresolved sources as discussed in section~\ref{sec:SPT}. 

The filtered mock SZE maps of the galaxy clusters give us the peak temperature of an unresolved source as a function of position within the cluster. We translate this into the flux using equation~(\ref{eqn:temp_to_flux}). The SZE flux extracted from the filtered mock maps is then used to boost the observed point source flux. This flux correction depends upon the position of the point source in the cluster as well as the mass and the redshift of the cluster.  
%A massive cluster has larger SZE flux, and in all cases the model flux decreases with distance from the cluster center.  As an example the SZE flux correction for two point sources which are $0.71^{\prime}$ and $2.2^{\prime}$ away from the center of the cluster (of mass $M_{500}$ of 6.83$\times10^{14}$ $\rm{M_{\odot}}$ and redshift 0.34) is 11.2 and 1.8 mJy, respectively at 150 GHz.  

As the angular size of a cluster decreases with redshift, a larger fraction of its SZE signature lies within a single SPT beam, and the angular distance of a cluster radio galaxy from the cluster center decreases.  For both these reasons the SZE flux bias correction tends to grow with redshift.  For example, two clusters with masses $M_{500}\sim4.3\times10^{14}\rm{M_{\odot}}$ that are at redshifts $z=0.075$ and $z=0.175$ have 150~GHz SZE flux corrections of 6.41~mJy and 8.61~mJy, respectively, for a point source residing at their centers. This flux is 1.2~percent and 7.3~percent of the total SZE flux in their $\theta_{500}$ regions.

It is important to note that the correction is only an approximation to the true flux bias, because there will in general be departures between the true (unobserved) Compton $Y$ profile of the cluster and the model we employ.  Nevertheless, because the true observed Compton $Y$ profiles of clusters have been shown to be in reasonably good agreement with the X-ray derived profiles \citep{plagge10}, this correction should be approximately correct in the mean as applied to an ensemble of cluster radio galaxies for a statistical study. We will discuss the systematics of this flux correction more in section~\ref{sec:flux_uncertainties}.

%The SZE flux correction depends on the cluster redshift (see Fig.~\ref{fig:flux_mass_redshift}). 
%As the angular size of the cluster decreases with redshift, the point sources in the cluster tend to lie closer to the cluster center in angular units.  In addition, each pixel in our maps covers a larger portion of the higher redshift cluster.  

%It is difficult to resolve a distant cluster, as the angular size of the cluster decreases with redshift. Thus a point source which is away from the center of the cluster tend to lie in the same resolution element as the center of the cluster. A clearly resolved cluster 

We note that as a result of the frequency dependent factor in equation~(\ref{eqn:temp_to_flux}), the SZE flux corrections at 150 GHz are 1.8 times larger than those at 95~GHz, for a fixed solid angle. However, the SZE flux corrections to the point sources at 95~GHz are found to be on average larger than the corrections at 150~GHz. This is due to the larger beam size at 95~GHz, which results in a 2.2 times larger effective solid angle for point sources ($\Delta \Omega_\mathrm{f}$ in equation~\ref{eqn:temp_to_flux}) at 95~GHz than at 150~GHz.

\subsubsection{SUMSS Based SPT 6~mJy Flux Limited Sample}
\label{sec:SPT_in_SUMSS}

\begin{figure}
\vskip-0.3in
\centering
\includegraphics[width=9cm]{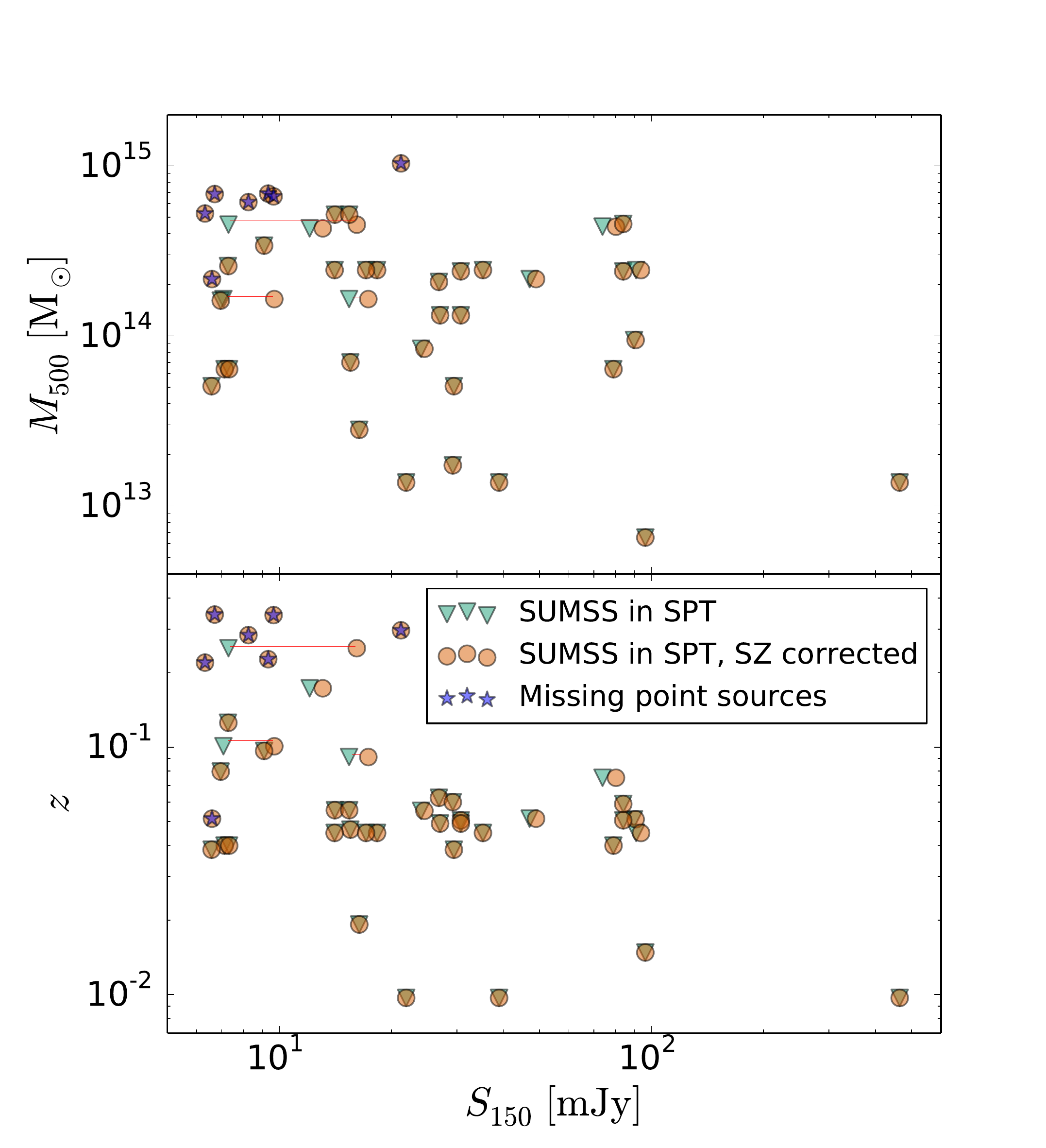}
\vskip-0.1in
\caption{SPT 150~GHz flux versus cluster redshift (bottom) and mass (top) for SUMSS selected radio galaxies with $S_{150}>6$~mJy that lie in the projected $\theta_{200}$ regions of clusters from the MCXC X-ray selected cluster catalog.  Green (brown) points show fluxes before (after) SZE flux corrections.  The SZE flux correction, in general, is larger for high mass and distant clusters compared to the low mass and nearby clusters, ranging between 0 and 24 mJy. Seven radio galaxies that would not have made the SPT flux cut because of the SZE flux bias from their host galaxy clusters are marked with stars.}
\label{fig:flux_mass_redshift}
\end{figure}

We focus on the locations of the SUMSS sources in the SPT 95, 150 and 220~GHz maps.  To build a flux limited radio galaxy catalog at SPT observing frequencies, we first check for a counterpart in the SPT detected point source sample using the method described in section~\ref{sec:sumss_plus_spt}.  Because both position and flux were determined simultaneously for the SPT detected point sources, the SPT fluxes must be corrected to account for the resulting flux bias as
\BE
S~ [{\rm mJy}] = \sigma_{\rm N} \sqrt{\xi^2 - 2},
\EE
where $\sigma_{\rm N}$ is the noise in the map and peak $\xi$ is the filtered $S/N$ measured at the SPT location of the source. If there is no counterpart in the SPT selected sample, we go to the CLEANed SPT maps and estimate the SPT flux at the position of the SUMSS source.  The CLEANed SPT maps are those where all sources with $S/N$ $(\xi)$ greater than 4.5 have been removed.  These maps are less affected by artifacts associated with bright point sources.  

In the SPT region, our sample contains 55,884 SUMSS sources above the completeness limits presented in section~\ref{sec:SUMSS}.  Above a flux limit of 6~mJy at 150~GHz, we find 2,970 sources in the SPT point source catalog, but after the flux debiasing we find 2,693 SPT counterparts of SUMSS sources above 6~mJy.   There are 37 sources above a flux limit of 6~mJy in the SPT detected catalog within the $\theta_{200}$ of the MCXC clusters.  In the SUMSS selected catalog there are 36 sources with flux above 6~mJy overlapping the MCXC clusters.  The SPT selected sample has a single extra point source, which lies just outside the cluster $\theta_{200}$ boundary according to its SUMSS position but just inside according to the SPT position.  Interestingly, this point source also has larger flux at 150~GHz as compared to 95~GHz and 843~MHz observing frequencies, indicating that it is possibly an SFG.  

We then apply the SZE flux corrections at 95 and 150~GHz for all the sources which are inside the MCXC clusters, using the method described in the last sub-section.  The number of candidate cluster radio galaxies increases to 43 after the SZE flux correction. Thus, we recover 7 additional sources, which were otherwise missing due to the SZE flux bias at 150~GHz.  Having a closer look at these sources, we find that all but one of these sources have negative fluxes before the SZE flux correction and are present in the central pixels of massive MCXC clusters, which are also counterparts of SPT confirmed galaxy clusters. One of the sources residing in a low redshift cluster has an SZE flux correction of around 4~mJy, which boosts it into the 6~mJy sample. These 43 SUMSS sources recovered in SPT maps at 150~GHz are shown in Fig.~\ref{fig:flux_mass_redshift} with the mass and redshift information of the host clusters. Each point with its uncorrected flux is shown in green and with its corrected flux in brown. The seven point sources recovered from the SUMSS catalog after the SZE flux correction are marked with a star, and in these cases we do not show the uncorrected flux.

\subsection{Radio Galaxy Spectral Indices}
\label{sec:alpha}
Following the technique in \citet{saro14}, we use a maximum likelihood analysis to estimate the spectral index ($\alpha$) of different samples of radio galaxies using different combinations of fluxes $S$ as
\BE
\label{eqn:alpha_lik}
{\cal L}(\alpha)
\propto \prod \limits_{i=1}^{N_\mathrm{source}} \mathrm{exp}\left({-{1\over2}\frac{
\left(S^{(i)}_{\nu_1} R(\alpha) - S^{(i)}_{\nu_2}\right)^2}{\left(\Delta
S^{(i)}_{\nu_1} R(\alpha)\right)^2 + \left(\Delta S^{(i)}_{\nu_2}\right)^2}}\right),
\EE
where $S^{(i)}_{\nu_1}$ ($\Delta S^{(i)}_{\nu_1}$) and $S^{(i)}_{\nu_2}$ ($\Delta S^{(i)}_{\nu_2}$) is the flux (flux uncertainty) of the $i^{\rm th}$ source at 0.843 and 95~GHz,  respectively, for the estimation of $\alpha_{0.843}^{95}$ (similarily for $\alpha_{0.843}^{150}$ and $\alpha_{95}^{150}$).  $R(\alpha)$ is given by
\BE
\label{eqn:R_alpha}
R(\alpha_{\nu_1}^{\nu_2}) = \left(\frac{\nu_2}{\nu_1} \right)^{\alpha_{\nu_1}^{\nu_2}}.
\EE

The most likely values of $\alpha_{0.843}^{95}$, $\alpha_{0.843}^{150}$ and $\alpha_{95}^{150}$ are listed in Table~\ref{tab:spectral_indices} for all SUMSS sources in the SPT-SZ region and for sources projected near the centers of the MCXC clusters, which we term BCGs (Brightest Cluster Galaxies). The SZE flux correction is applied to the source fluxes at 95 and 150~GHz.  There is a tendency for the spectra to steepen with frequency, and the central sources (BCGs) tend to exhibit steeper spectra than the typical radio galaxies in the field.

Similar spectral indices (within our quoted error bars for central sources/BCGs in MCXC clusters) were found by \citet{lin07} where they used the NRAO VLA Sky Survey \citep[NVSS,][]{condon98} 1.4~GHz data together with the 4.85~GHz data from the GB6 \citep[Green Bank 6 cm survey,][]{gregory96} and PMN \citep[Parkes-MIT-NRAO survey,][]{griffith93}.  They reported $\alpha_{1.4}^{4.85}=-0.51$ for most of the sources in the cluster,  and they also found steeper spectra for BCGs in their sample. Our results are also consistent with other previous analyses \citep{condon92,cooray98,coble07}. 
%A recent work by \citet{sayers13a} reported  a median value of $\alpha_{1.4}^{30}=-0.89$ in a set of 28 cluster member galaxies.

%
\begin{table}
\footnotesize
\caption{The characteristic spectral indices and 1$\sigma$ uncertainties for all SUMSS detected sources in the SPT region and a subset of these, which we denote as BCGs, that lie within $0.1\times\theta_{200}$ of the MCXC cluster centers. Mean spectral indices are presented for pairs of frequencies constructed from 150~GHz, 95~GHz and 843~MHz.  The SZE correction is applied at 95 and 150~GHz.}
\label{tab:spectral_indices}
\begin{center}
\begin{tabular}{lccc}
\hline
Dataset & $\alpha_{0.843}^{150}$ & $\alpha_{0.843}^{95}$ & $\alpha_{95}^{150}$\\[3pt]
\hline
%SPT detected    & $-0.13^{+0.43}_{-0.34}$	& $-0.12^{+0.35}_{-0.39}$ &$-0.52^{+0.31}_{-0.32}$ \\[3pt]
SUMSS & $-0.38^{+0.28}_{-0.29}$	& $-0.38^{+0.28}_{-0.31}$ & $-0.50^{+0.24}_{-0.23}$ \\[3pt]
SUMSS BCGs  & $-0.63^{+0.34}_{-0.29}$ & $-0.64^{+0.33}_{-0.40}$ & $-0.77^{+0.32}_{-0.31}$ \\[3pt]
\hline
\end{tabular}
\end{center}
\end{table}

\section{Results}
\label{sec:results}

In this section we use the flux limited samples of radio point sources to first study the radial profile of high frequency cluster radio galaxies and to then study the LF.  A radial profile is needed for the deprojection of the measured LF into the cluster virial volume.

\begin{figure}
\vskip-0.2in
\centering
\includegraphics[width=9cm]{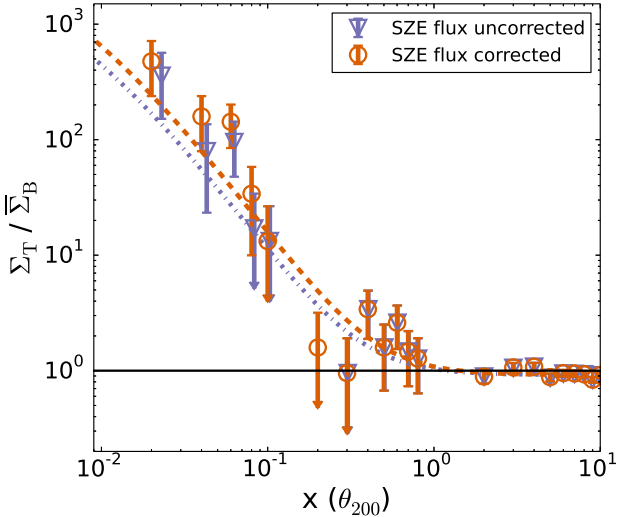}
\vskip-0.1in
\caption{Surface density profiles for two flux limited $S_{150}>6$~mJy samples of radio galaxies stacked within 139 MCXC clusters.  Both samples with SZE flux corrections (red) and without (blue) are shown.  The lines are the best fit NFW models (see Table~\ref{Table:NFWparameters}).}
\label{fig:SD_profiles}
\end{figure}

\subsection{Radial Distribution of Cluster Radio Galaxies}
\label{sec:radial_profiles}
We examine the distribution of radio galaxies in the cluster virial region by stacking all radio galaxies overlapping the MCXC sample in the coordinate $\theta/\theta_{200}$. We use the projected NFW profile \citep{navarro97} as a fitting function for the radial distribution, where the projected profile can be written as \citep{bartelmann96}:
\BE
\Sigma(x) = {2\rho_{\rm s}r_{\rm s}\over x^2 - 1}f(x),
\label{eqn:sigma_of_x1}
\EE
with $\rho_{\rm s}=\rho_{\rm c}\delta_{\rm c}$ (where $\rho_{\rm c}$ is the critical density of the Universe and $\delta_{\rm c}$ is a characteristic density contrast), $r_{\rm s}$  is the typical profile scale radius and $f(x)$ is given by
\begin{displaymath}
f(x) =
\begin{cases}
  1- \frac{2}{\sqrt{x^2-1}}{\rm arctan}\sqrt{\frac{x-1}{x+1}} & \text{if } x > 1, \\
  1- \frac{2}{\sqrt{1-x^2}}{\rm arctanh}\sqrt{\frac{1-x}{x+1}} & \text{if } x < 1,\\
  1/3 & \text{if } x = 1.%\\
\end{cases}
\end{displaymath}
Here $x=r/r_{\rm s}$ and $r_{\rm s}=R_{200}/c$, where $c$ is the concentration parameter. Following \citet{lin04a}, we remodel the projected NFW profile by integrating over equation~(\ref{eqn:sigma_of_x1}) and get the projected number of galaxies
\BE
N(x)=\frac{N_{200}}{g(c)}g(x),
\label{eqn:N_of_x}
\EE
where the normalization $N_{200}$ is the number of galaxies projected in the cluster virial radius $R_{200}$ and $g(x)$ is given as
\BE
g(x) = \int_{0}^{x} \frac{x^\prime f(x^\prime)}{(x^{\prime 2}-1)}{\rm d}x^\prime,
\EE
where $x$ is equivalent to $c$ for $r=R_{200}$ to give $g(c)$.
This reduces the covariance between the normalization and concentration parameters of the NFW profile.

The surface density of the clusters can have both cluster and background components and is written as
\BE
\Sigma_{\rm T}=\Sigma(x)+\Sigma_{\rm B},
\EE 
and in terms of the number of galaxies as
\BE
N_{\rm T}=N(x)+\Sigma_{\rm B}A,
\label{eqn:no_gal}
\EE 
where $A$ is the solid angle of the annulus or bin. Thus we fit our stacked distribution of radio galaxies to a model with three parameters: $c$, $N_{200}$ and $\Sigma_{\rm B}$. We stack radio galaxies out to $10\times\theta_{200}$ to allow for a good constraint on the background density $\Sigma_{\rm B}$.

We employ the \citet{cash79} statistic 
\BE
\label{eqn:cash_lik}
\it{C} = \sum_{i} \left(
N_{{\rm T},i}^{\rm d} \ln(N_{{\rm T},i}^{\rm m}) - N_{{\rm T},i}^{\rm m} - N_{{\rm T},i}^{\rm d} \ln(N_{{\rm T},i}^{\rm d}) + N_{{\rm T},i}^{\rm d}
\right),
\EE
in this fit, where $N_{\rm T, i}^{\rm m}$ is the total number of galaxies from the model as in equation~(\ref{eqn:no_gal}) and $N_{\rm T, i}^{\rm d}$ is the total number of galaxies in the observed data in the $i^{\rm th}$ angular bin.  This is just the difference in $N(x)$ evaluated at the outer and inner boundaries of each bin. 

We use the Markov Chain Monte Carlo (MCMC) code, {\tt emcee} \citep[a Python implementation of an affine invariant ensemble sampler;][]{mackey13} to fit the model to the data for two different datasets:  (1) the SUMSS based sample of radio galaxies with uncorrected fluxes, and (2) the sample with SZE flux corrections.  In the fitting we adopt a bin size corresponding to $\theta_{200}/1000$ and fit over the region extending to $10\theta_{200}$. 
%the $10\theta_{200}$ region we study, there are a total of 2105 and 1970 radio galaxies in the SPT selected and the SUMSS based samples.
The concentration parameter is sampled in $\rm log$ space during the fit.

The best fit values and uncertainties of the parameters for different datasets are given in Table~\ref{Table:NFWparameters}.  The background subtracted number of galaxies $N_{200}$ in the stack of 139 galaxy clusters is $\sim30$ ($\sim20$) in the SZE flux corrected (uncorrected) sample.   There is an increase in the normalization $N_{200}$ in going from the sample with uncorrected fluxes to the SZE bias corrected sample of approximately 50~percent, because the SZE flux bias correction only affects sources lying projected onto the cluster virial region, and the additional sources that come into the flux limited sample are predominantly cluster radio galaxies.

$\Sigma_{\rm B}$ is the background density $[\mathrm{deg}^{-2}]$ which can be multiplied by the total solid angle ($\simeq20.5$~deg$^2$) of the cluster stack within $\theta_{200}$ to get an estimate of the number of background galaxies. $\overline{N}_{\rm T}$ is the total number of galaxies within $\theta_{200}$ of the stacked clusters above our flux limit of 6~mJy, and this is close to the sum of $N_{200}$ and the number of background galaxies obtained from the fit.  Because $N_{200}$ is evaluated from a stack of clusters, we have between 0.15 and 0.20 radio galaxies per cluster.  If we sum the virial masses of the MCXC clusters we then have between 0.5 and 0.75 radio galaxies per $10^{15}\rm M_\odot$ of cluster mass.  The profile is strongly centrally concentrated with $c\sim100$, indicating that the radial distribution of cluster radio galaxies is consistent with a power law distribution $n(r)\propto r^{-3}$.  We use this behavior in the next section to correct the projected LF to the LF within the cluster virial region defined by $r_{200}$.

In Fig.~\ref{fig:SD_profiles} we show the best fit surface density profiles and data for each of the two datasets. To create these plots we combined many bins to reduce the noise in the measured radial profile. We normalize the $y$-axes of this plot with the annulus area in each angular bin and the background level density or the mean number density of sources ($\overline{\Sigma}_{\rm B}$) in the SPT region so that we can compare the surface density profiles from the two datasets. 
It is worth noting that $\overline{\Sigma}_{\rm B}$ is not same as the background density ($\Sigma_{\rm B}$), where the latter is one of the fit parameters. However, Fig.~\ref{fig:SD_profiles} shows that the mean survey density is a good estimation of the background number density of the clusters, as $\Sigma_{\rm T}$/$\overline{\Sigma}_{\rm B}$ is consistent with 1 outside of the cluster. 

\begin{table}
\caption{Best fit NFW model parameters for the radial profile of radio galaxies with $S_{150}\ge6$~mJy in a stack of 139 MCXC clusters. The samples with uncorrected and SZE corrected fluxes are shown, and for each we present concentration $c$, $N_{200}$, background density $\Sigma_{\rm B}$, and the number of sources $\overline{N}_{\rm T}$ within $\theta_{200}$.}
\label{Table:NFWparameters}
\begin{center}
\begin{tabular}{lcccc}
\hline
\multicolumn{1}{l}{Dataset} & \multicolumn{1}{c}{$c$} & \multicolumn{1}{c}{$N_{200}$} & \multicolumn{1}{c}{$\Sigma_{\rm B}$ [$\rm deg^{-2}$]} & \multicolumn{1}{c}{$\overline{N}_{\rm T}$}\\
\hline
Flux uncorrected    & $107^{+277}_{-51}$	     & $19.7^{+5.7}_{-4.8}$       & $0.94\pm0.02$ &   36 \\[3pt]
Flux corrected    & $108^{+107}_{-48}$       & $28.7^{+6.2}_{-5.6}$       & $0.94\pm0.02$ &  43 \\
\hline
\end{tabular}
\end{center}
\end{table} 

\subsection{Cluster Radio Galaxy Luminosity Functions }
\label{sec:LFs}
In this section, we construct radio LFs using the excess of radio sources toward galaxy clusters and assigning those excess sources to the cluster redshift \citep[following][]{lin04a}.  We calculate the LFs not only for the SPT bands but also for the SUMSS band.  We compute the radio luminosity of the SUMSS point sources overlapping the MCXC galaxy clusters using the observed fluxes (before and after SZE correction) and the redshift of the respective cluster. In the luminosity calculations we apply the redshift dependent $\it k$-correction in an attempt to estimate the luminosity at the same rest frame frequency for all redshifts. Thus the radio source luminosity is given by:
\BE
\label{eqn:luminosity}
P_{\nu_{\rm S}} \ = \ (4\pi \ D_{\rm L}^2) \ S_{\nu_{\rm S}} \ \frac{ {\it k}(z) } { (1+z) },
\EE
where $D_{\rm L}$ is the luminosity distance to the redshift $z$ of the cluster, $S_{\nu_{\rm S}}$ is radio source flux at frequency $\nu_{\rm S}$ and ${\it k}(z)$ is the $\it k$-correction given by (1+$z$)$^{-\alpha}$. We choose a spectral index $\alpha=-0.8$ for the 843~MHz analysis and an $\alpha=-0.5$ for the higher frequency analyses (see Table~\ref{tab:spectral_indices} for results at higher frequency). 

\subsubsection{LF Fitting Method}
\label{sec:LFmethod}
To construct the LF we consider all the MCXC clusters that lie in the SPT (or SUMSS) region, adding up the number of point sources within $\theta_{200}$ in different logarithmic luminosity bins (effectively placing all radio galaxies at the redshift of the cluster).  For each luminosity bin, we estimate the background counts from the population of observed sources in the SPT (or the SUMSS) region, in the corresponding bins in $\rm log$$N$$-$$\rm log$$S$ space, where we use the cluster redshift to transform from radio galaxy flux to luminosity. These background counts are corrected for the surface area of all the clusters in our sample. We also keep track of the total mass $\Sigma M_{200}$ of the clusters, which are contributing to each of the luminosity bins.  We use this vector of total masses to normalize our LF, allowing us to account for the fact that with a particular flux limit the high redshift cluster radio galaxies do not extend to as faint a luminosity as those in the low redshift clusters.  Another way is to normalize it with the total volume of these clusters. However, doing so introduces a redshift dependence in the LF as we define the virial region $\theta_{200}$ as the region with an overdensity of 200 times the critical density of the universe at that redshift, and the critical density scales as $E^2(z)$.  Thus, normalizing by total mass is a good choice, because if the AGN activity were independent of redshift we would expect to see the same LF defined as the number of galaxies per unit mass at all redshifts.  In addition, this normalization facilitates comparison of the field and cluster LFs to determine whether AGN activity depends on environment.

\begin{figure}
\vskip-0.2in
\centering
\includegraphics[width=9cm]{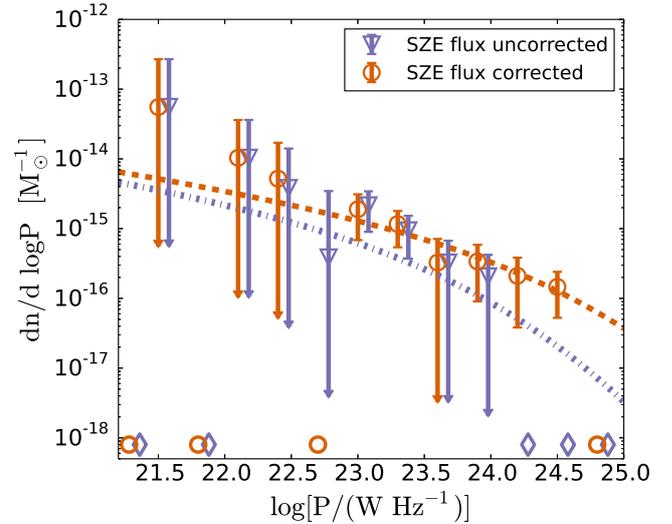}
\vskip-0.1in
\caption{The 150~GHz LF for sources within $\theta_{200}$ of the centers of massive galaxy clusters. This LF is derived from the SUMSS selected sample in the SPT region with (red) and without (blue) SZE flux corrections. Lines are the best fitting LF models. The increase in LF amplitude on the high luminosity end when using the SZE flux bias corrected sample is clear.  For convenience in this figure the bins containing negative values in the background subtracted counts are represented as points at the bottom of the figure.}
\label{fig:LF_different_spt_sumss}
\end{figure}

To fit the LF we again use MCMC analysis with the Cash statistic as described in equation~\ref{eqn:cash_lik}.  We first attempt to fit a Schechter function \citep{schechter76}, but this is a poor fit to the data.  Thus we take the functional form used in \citet{condon02} for our fits. The LF model is 
\BE
\label{eqn:Condon_fit}
\log \left( \frac{{\rm d}n}{{\rm d} \log P} \right) = y - \left[ b^2 + \left( \frac{\log P-x}{w} \right)^2 \right]^{1/2} - 1.5 \log P,
\EE
where the parameters $b$, $x$ and $w$, control the shape of the LF and $y$ is its amplitude.
The LF shape parameters are determined in \citet{condon02} for AGN and star forming galaxies (SFGs) at low frequency and for the field population. The shape parameters are ($b_1$, $x_1$, $w_1$) = (2.4, 25.8, 0.78) for the AGN and ($b_2$, $x_2$, $w_2$) = (1.9, 22.35, 0.67) for the SFGs.  To evaluate the likelihood of a given model, we take the LF model and scale by the total mass of the sample of clusters contributing to each luminosity bin and then add the background number of galaxies determined from the data for that bin.  That is, we do not fit to the background subtracted counts.  We validate our code by analyzing simulated samples created using the best fit LFs reported below, demonstrating that we recover the input parameters.

We scale the LF amplitude to account for cluster radio galaxies projected onto the virial cylinder but not lying in the virial sphere;  this deprojection correction $D_{\rm prj}$ has a very small impact for the radio galaxy case, because the radial distributions are so centrally concentrated.  Specifically, $D_{\rm prj}=0.92$ for an NFW concentration of 108, which is the best fit value listed in Table~\ref{Table:NFWparameters}. The 2-$\sigma$ excursion from the mean concentration to lower (34) and upper (460) values correspond to deprojection values of 0.9 and 0.94, respectively, and thus the uncertainty on the concentration does not impact our LF measurements significantly.

Following \citet[][]{lin07}, we first fit the sum of the AGN and SFGs \citet{condon02} models to the SUMSS data above the completeness limits at 843~MHz by allowing the amplitudes ($y_1$ and $y_2$) and x-axis scales ($x_1$ and $x_2$) to vary, while fixing the other shape parameters of the function.  
%We find that the shape parameter $x_2$ for SFGs does not converge in our fit, and we only find an upper limit for the amplitude $y_2$. In contrast, we find good estimates of the best fit values and uncertainties for the $y_1$ and $x_1$ parameters that describe the AGN population. 
We find that SFG population is not large enough to get meaningful constraints on the SFG part of the function. This is expected, because at the SUMSS  depths and frequency 843~MHz we are probing well the more luminous AGN population but not the fainter SFG population.  In addition, in clusters we would expect the SFG population to be suppressed, making it even harder to constrain.  Thus, we fit just for the AGN part of the LF by varying $x_1$ and $y_1$ parameters in the MCMC chain \citep[while keeping other shape parameters for the AGN part of the LF fixed to][]{condon02}. We adopt this fitting approach of ignoring the SFG contribution also for the high frequency LFs.

We also validate our fitting code using a much larger sample of radio-loud AGN to construct the field LF \citep{best12}.  Fitting their dataset (see table~2 of their paper) using the LF described in equation~(\ref{eqn:Condon_fit}), we find ($y_1$, $b_1$, $x_1$, $w_1$) = ($33.79^{+0.51}_{-0.37}$, $1.88^{+0.5}_{-0.4}$, $25.48^{+0.08}_{-0.07}$, $0.74^{+0.04}_{-0.04}$), in good agreement with \citet{condon02}.  We see only small differences in our results if we keep $b_1$ and $w_1$ fixed to either \citet{condon02} or \cite{best12} values.  Thus, we see no sensitivity of our fitting parameters to the decision of whether to adopt \citet{condon02} or \citet{best12} shape parameters.

\begin{figure*}
\vskip-0.2in
\centering
\includegraphics[width=8.8cm, height=7.75cm]{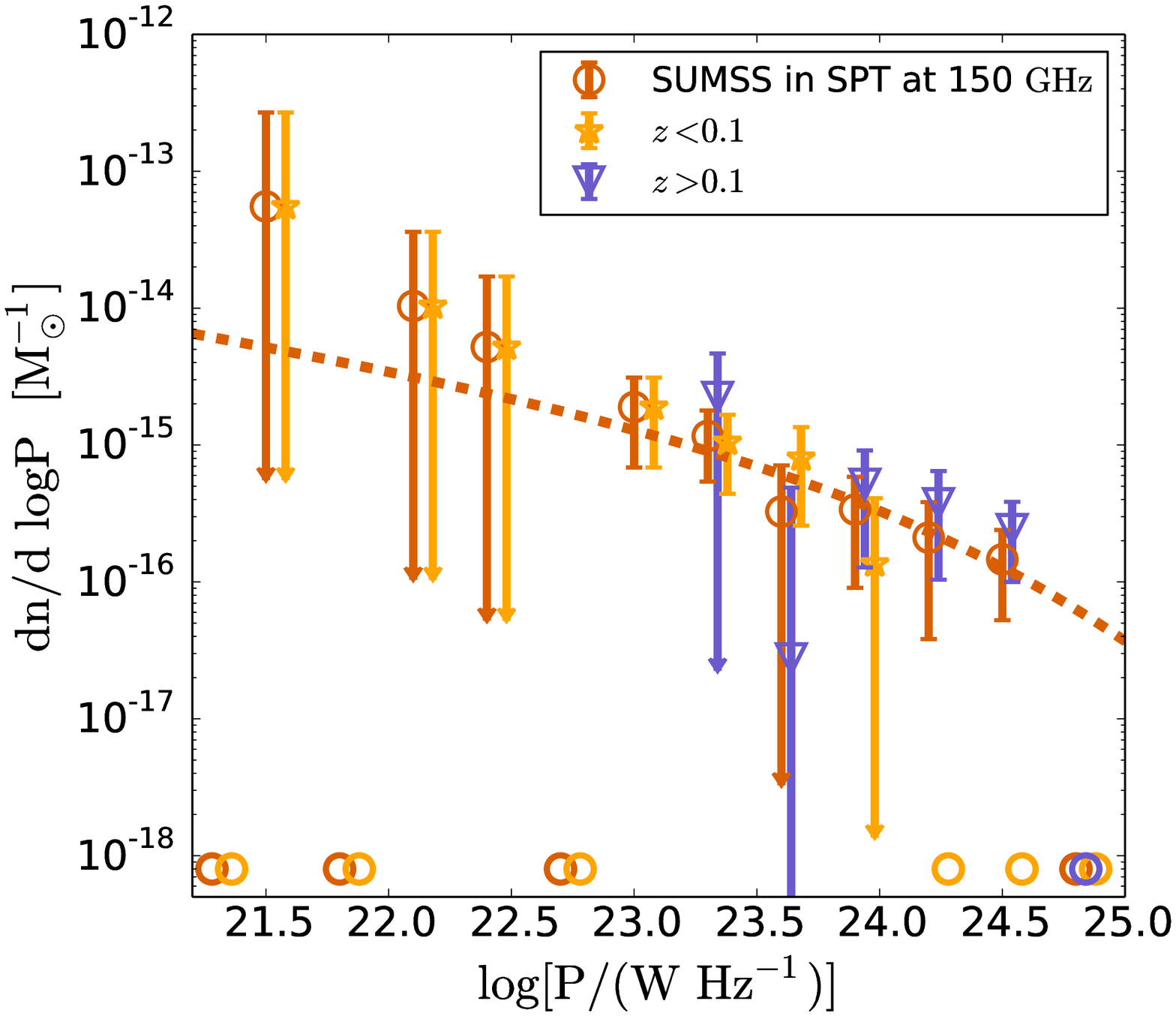}
\includegraphics[width=8.8cm, height=7.75cm]{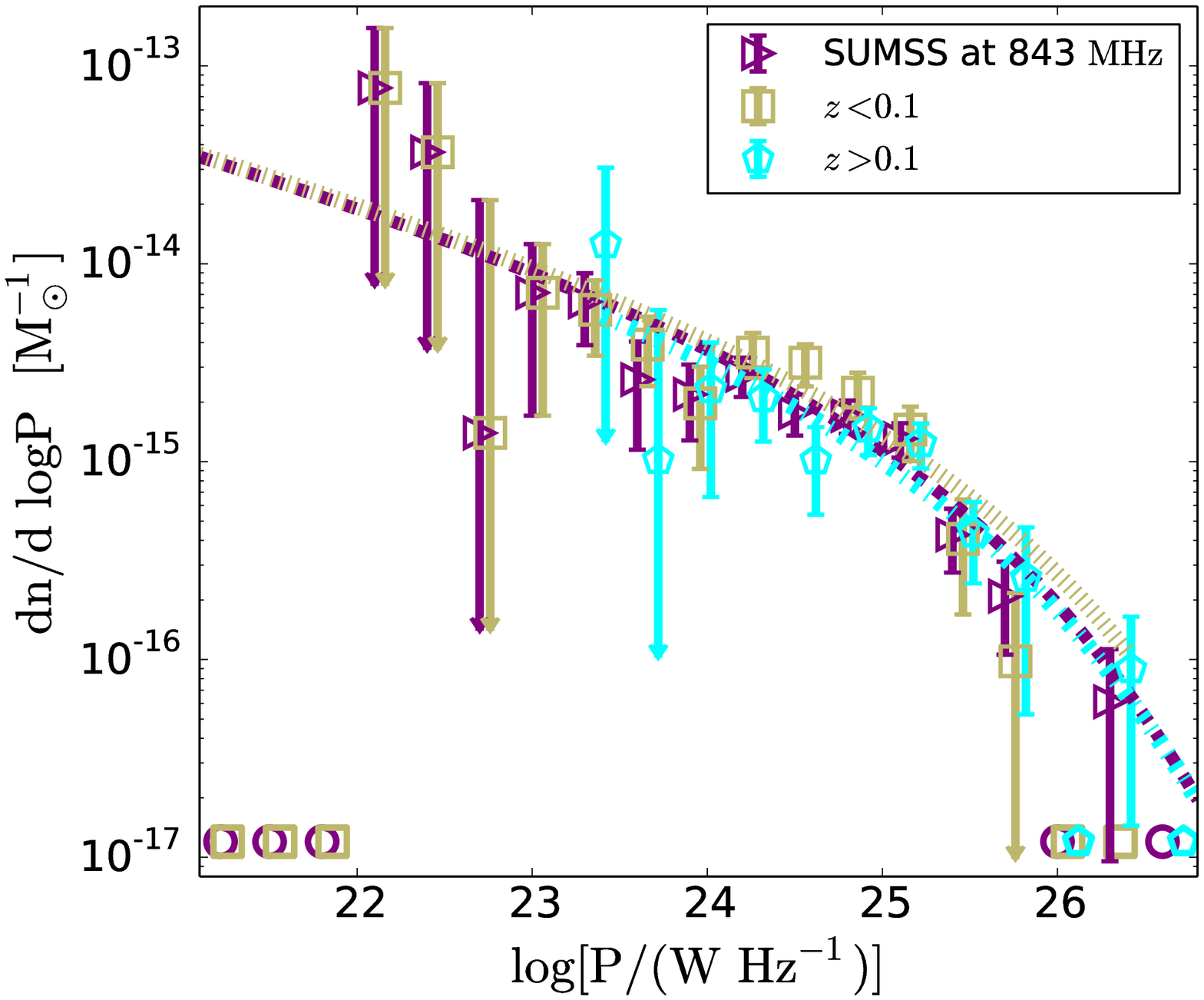}
\vskip-0.1in
\caption{Cluster Radio Galaxy LFs: The SUMSS based 150~GHz LF (left), which includes SZE flux bias corrections, and the SUMSS 843~MHz LF (right), which is constructed using MCXC clusters over the full 8100~deg$^2$ SUMSS survey region. The datasets are fitted with the AGN component of the LF by varying $y_1$ and $x_1$ parameters as discussed in section~\ref{sec:LFmethod}. The data points are shifted horizontally to improve visibility. Different lines indicate the best fit model LFs (see Table~\ref{tab:LFparameters}). In both plots, we divide the samples into two different redshift bins. However, the data are not enough to provide meaningful constraints on the redshift evolution for the 150~GHz LF. For convenience in this figure the bins containing negative values in the background subtracted counts are represented as points at the bottom of the figure.}
\label{fig:L_function}
\end{figure*}
\begin{table}
\footnotesize
\caption{The best fit LF parameters for different samples of cluster radio galaxies. The samples of SPT fluxes at SUMSS locations ``SUMSS in SPT") are corrected for the SZE flux bias at 95 and 150 GHz, except for ``SUMSS in SPT (U)", which denotes the sample with uncorrected fluxes.}
\label{tab:LFparameters}
\begin{center}
\begin{tabular}{l c c c}
\hline
Dataset & $\nu$ $(\rm GHz)$ & $y_1$ & $x_1$ \\[5pt]
\hline
SUMSS                    &  0.843    & $25.90^{+0.19}_{-0.18}$   & $26.81^{+0.20}_{-0.18}$ \\[5pt] 
$z< 0.1$                  & 0.843     & $26.10^{+0.40}_{-0.31}$    & $27.02^{+0.38}_{-0.30}$ \\[5pt] 
$z> 0.1$                  & 0.843     & $25.88^{+0.28}_{-0.27}$    & $26.86^{+0.30}_{-0.29}$ \\[5pt] 
SUMSS in SPT       & 95          & $23.89^{+0.46}_{-0.37}$    & $25.57^{+0.51}_{-0.44}$ \\[5pt]
SUMSS in SPT (U) & 150        & $22.47^{+0.70}_{-1.62}$    & $24.62^{+2.53}_{-0.89}$ \\[5pt] 
SUMSS in SPT       & 150        &$23.46^{+0.62}_{-0.46}$     & $25.34^{+0.74}_{-0.57}$ \\[5pt]
SUMSS in SPT       & 220        &$22.58^{+0.33}_{-1.06}$    & $24.27^{+1.06}_{-0.77}$ \\[5pt]
\hline
\end{tabular}
\end{center}
\end{table}

\subsubsection{LF Measurements}

The 150~GHz LFs are shown in Fig.~\ref{fig:LF_different_spt_sumss} for the SUMSS based sample of radio galaxies with uncorrected fluxes, and the sample with SZE flux corrections.  In this figure as in all other LF figures, we show the background subtracted observed counts binned within much larger luminosity bins to improve the signal to noise.  These figures do not properly represent the LF fitting method described above, but are convenient for showing comparisons of data and best fit models.

The LF has higher amplitude at high luminosities after the SZE flux correction. This increase is due to the additional sources that come into the sample once the bias corrections are applied. While these are low flux sources, the SZE corrections are larger at higher redshifts, so they have a larger impact on the high luminosity radio galaxy population.  In Table~\ref{tab:LFparameters} the datasets are listed in the first column followed by the frequency of the sample and then the two LF parameters $y1$ and $x1$. The best-fit parameters for the 150~GHz luminosity function before and after SZE correction are different, but given the uncertainties, the differences are not statistically significant. 

Next we construct the LF of SUMSS sources at 843~MHz within the $\theta_{200}$ of MCXC clusters as shown in the right panel of Fig.~\ref{fig:L_function}. Because the SZE flux is negligible at 843~MHz, no correction is required in the flux measurements of the SUMSS sources. We choose the flux cut at the 100~percent completeness limits of the SUMSS catalog described in section~\ref{sec:SUMSS}.

We probe for changes with redshift by measuring the LFs in two different redshift bins.  To do this, we separate our MCXC cluster population into two redshift bins having similar numbers of galaxy clusters.  Given the low redshift nature of the MCXC sample we split at redshift $z=0.1$.  In the SUMSS region, we divide the cluster sample into two parts with 159 (174) clusters over 8100~$\deg^2$ at $z \leq0.1$ ($z \geq0.1$), and construct the LF for these samples (see the right panel of Fig.~\ref{fig:L_function}).  We see no evidence of redshift evolution of the LF, and indeed the measurements that we list in Table~\ref{tab:LFparameters} reflect this lack of evolution.

For the SUMSS based sources with fluxes measured in the SPT maps, the low luminosity end of the LF at 150~GHz mainly consists of galaxy clusters stacked at $z\leq$0.1, as shown in the left panel of Fig.~\ref{fig:L_function}.  At this frequency, there are only 10 SUMSS detected sources within the $\theta_{200}$ of the galaxy clusters above $z\geq0.1$, and 6 of them are there because of the SZE flux correction.  Thus, there are not enough data to constrain the redshift evolution, but certainly in Fig.~\ref{fig:L_function} the two subsamples do not appear to be different.  We do not present the best fit parameters of the two fits in Table~\ref{tab:LFparameters}.

\begin{figure}
\vskip-0.2in
\centering
\includegraphics[width=9cm]{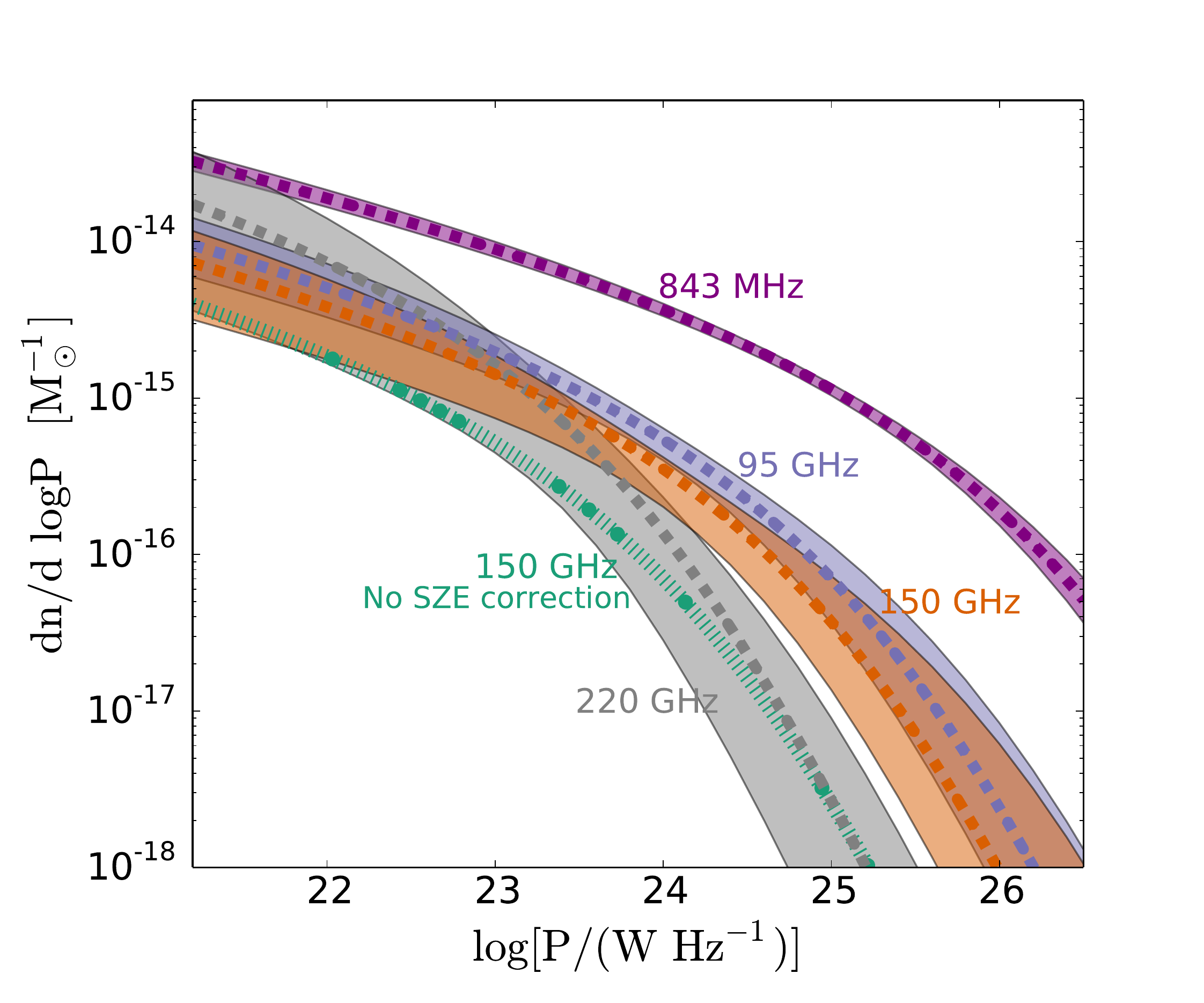}
\vskip-0.1in
\caption{LF fits to samples at different observing frequencies. As explained in section~\ref{sec:LFmethod} the fit is done using an AGN fitting function \citep{condon02}.  The filled regions show the best fit model and the $1\sigma$ confidence regions (see Table~\ref{tab:LFparameters}).  The curves show the decrease in the cluster radio galaxy population with increasing frequency and increasing power, and -- in the 150~GHz case -- the impact of the SZE flux bias correction.}
\label{fig:LF_different_neu}
\end{figure}

Finally in Fig.~\ref{fig:LF_different_neu} we plot the best fit LFs at 0.843, 95, 150 and 220~GHz observing frequencies. For comparison we include the 150~GHz LF before SZE flux bias corrections are applied, whereas for 95~GHz only the SZE flux bias corrected LF is shown.  As discussed earlier, we fit these LFs using only an AGN component (equation~\ref{eqn:Condon_fit}) with varying $x1$ and $y1$ with the other parameters fixed.  These LFs are constructed using data that are 100~percent complete at 843~MHz and with a flux cut of 6~mJy at 95, 150 and 220~GHz frequencies to enable a comparison of the radio galaxy populations to the same flux limit.   The number of candidate sources in clusters at 95, 150 and 220~GHz is 65 (34.7), 43 (22.1) and 64 (16.8) before (after) background subtraction, respectively. These numbers are small, and therefore it is not possible to make precise comparisons between the LFs.  Nevertheless, it is evident from this plot that the amplitude of the 843~MHz LF is approximately one order of magnitude higher than the amplitude of the high frequency LFs.  We show the 150~GHz LF before and after the SZE correction, indicating the significance of accounting for the cluster SZE bias at this frequency.  The best fit parameters for the \citet{condon02} fitting function are given in Table~\ref{tab:LFparameters} at different frequencies and for the different datasets.

%The LFs at 95, 150 and 220~GHz are interesting to compare. Given the limited size of our high frequency cluster radio galaxy sample, the differences among these LFs shown in Fig.~\ref{fig:LF_different_neu} have low statistical significance.  Generally speaking, all of these LFs have similar amplitude at low power, but the 220~GHz LF falls off more steeply than the 95~GHz LF.  The 150~GHz LF shows intermediate behavior.  The high frequency radio galaxies are approximately an order of magnitude rarer than those at 843~MHz.   

%%%%%%%%%%%%%%%%%%%%%%%%%%%%%%%%%%%%%%%%%%%%%
%%%%%%                        Impact on SZE Cluster Selection                        %%%%%%%%
%%%%%%%%%%%%%%%%%%%%%%%%%%%%%%%%%%%%%%%%%%%%%

\section{Radio Galaxy Contamination of Cluster SZE}
\label{sec:contamination}
The LFs presented in the last section describe the number of radio galaxies inside a galaxy cluster of a given mass and redshift. The collective flux of these cluster radio galaxies can contaminate the SZE signature of a galaxy cluster, potentially affecting the observability of the cluster and the accuracy of the derived virial mass estimate. To quantify these effects for an SPT-SZ like SZE cluster survey we use the LFs derived from the SUMSS based measurements at 95 and 150 GHz in the SPT maps.  These measurements include the SZE flux bias corrections and therefore are our best available estimates of the true underlying 95 and 150~GHz cluster radio galaxy LFs over the mass and redshift ranges of the MCXC sample. 

\begin{figure*}
\vskip-0.2in
\centering
\includegraphics[width=9.4cm, height=7.7cm]{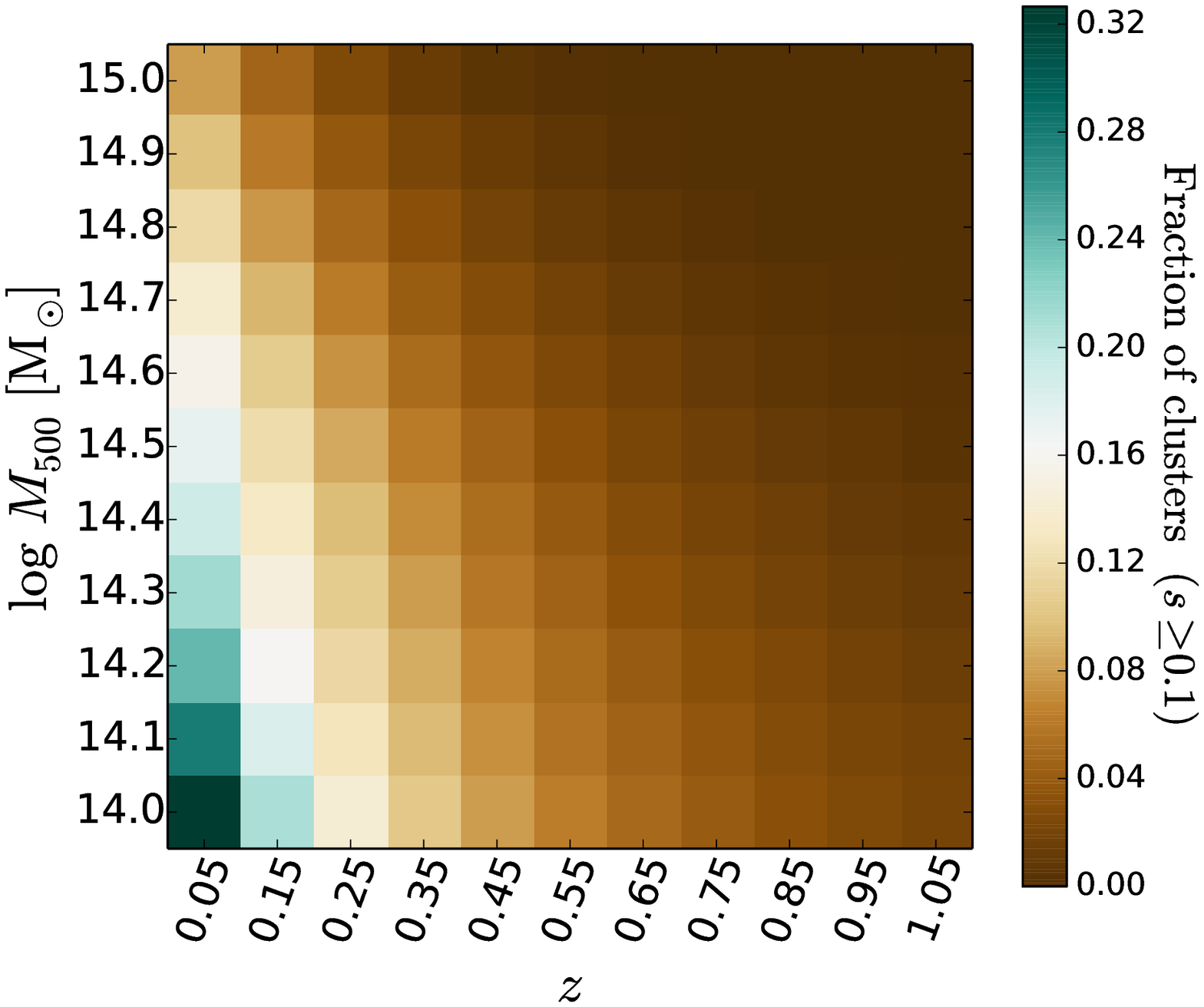}
\includegraphics[width=8.2cm, height=7.7cm]{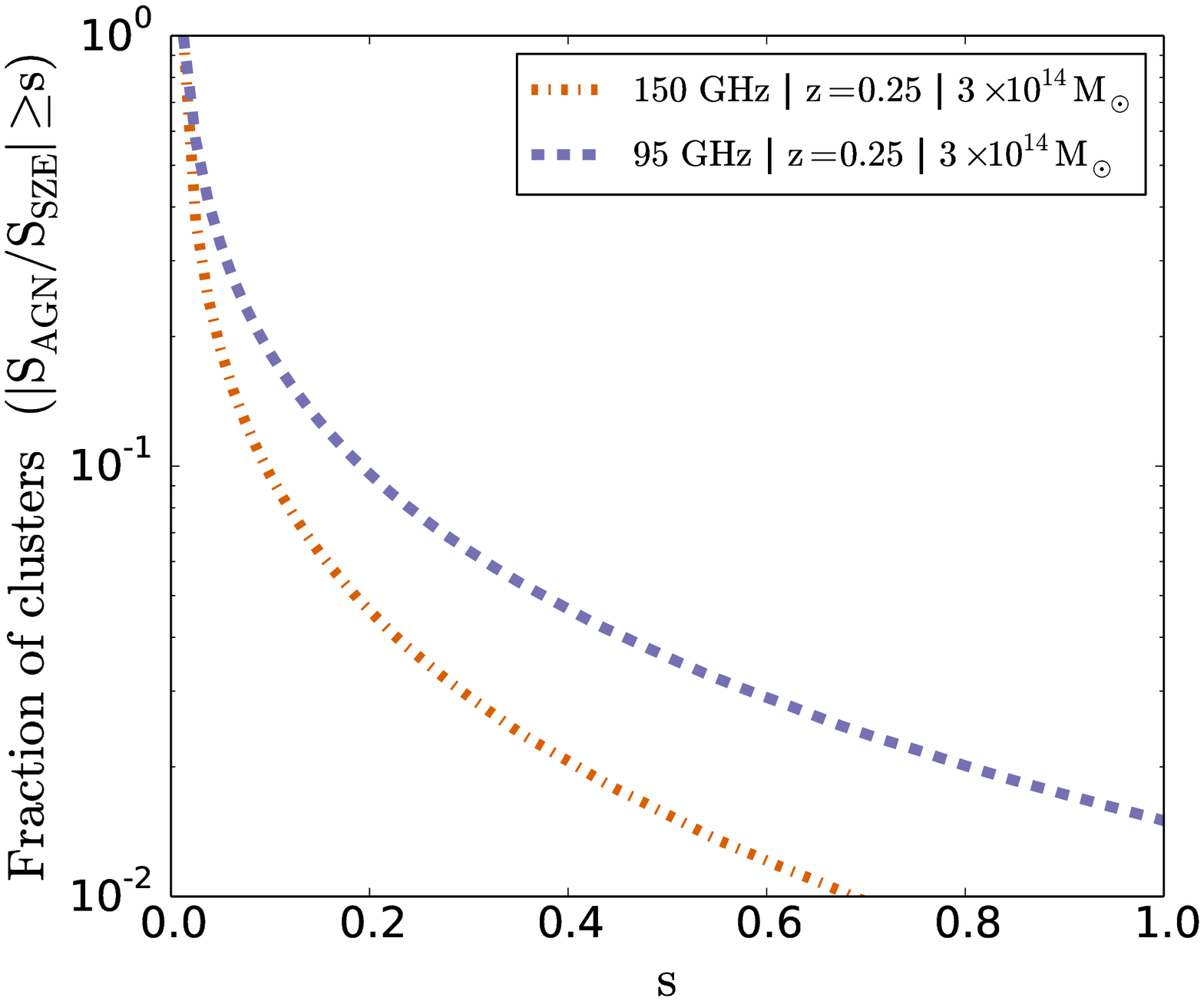}
\vskip-0.1in
\caption{Fraction of clusters contaminated above a degree of contamination $s\geq 0.1$ for clusters as a function of mass $M_{500}$ and redshift at 150 GHz (left panel). This shows that there is a higher probability of missing the detection of low redshift and low mass clusters. Right panel shows the fraction of clusters (with $M_{500}=3\times 10^{14} \rm M_{\odot}$ and $z=0.25$) above a given degree of contamination $s$ at 95 and 150~GHz. The contamination is larger at 95 $\rm GHz$, both because of the smaller SZE signature and the higher AGN fluxes at this frequency as compared to that at 150~GHz.}
\label{fig:Contamination1}
\end{figure*}

\subsection{Characteristic Levels of Contamination}

To estimate the cluster population of radio galaxies, we multiply the LF with the mass of the cluster of interest and integrate it in the luminosity range of 10$^{21}$ to 10$^{27}$ $\rm WHz^{-1}$, producing an expectation value $<N_{\rm A}>$ for the expected number of cluster radio galaxies in this luminosity range.   

To model the effects of radio galaxies in a way that accounts for the cluster to cluster variations of the population, we randomly sample a Poisson distribution with mean $<N_{\rm A}>$ to determine the number $N_{\rm R}$ of radio galaxies for a cluster of particular mass and redshift. For each radio galaxy we then assign a flux using the LF at that frequency as the probability distribution function and then sum the fluxes from the $N_{\rm R}$ cluster radio galaxies to get the total contaminating cluster radio galaxy flux $S_{\rm A}$ in that cluster. We define the degree of contamination to be
\BE
\label{eqn:s_contamination}
s=\left|{S_{\rm A}\over S_{\rm SZE}}\right|,
\EE
where the total cluster radio galaxy flux is $S_{\rm A}$ and the total (negative) cluster SZE flux within $\theta_{200}$ is $S_{\rm SZE}$.  The SZE flux of the cluster is derived from the integrated  $Y_{\rm SZ}$ parameter, using the pressure profiles as described in \citet{arnaud10} and is converted into the same units as the cluster radio galaxy flux
\BE
\label{eqn:SZE_flux}
S_{\rm SZE} = g_{\nu} f_{\nu} I_0 Y_{\rm SZE} ,
\EE
where $I_0$ is equivalent to $2(K_{\rm B} T_{\rm CMB})^3 / (hc)^2\simeq 2.7033\times 10^8$ $\rm Jy/sr$, $Y_{\rm SZ}$ has units of steradian, $g_{\nu}$ and $f_{\nu}$ give the frequency dependence of the survey such that
\BE
g_{\nu}=x\coth\left(\frac{x}{2}\right)-4,
\EE
\BE
f_{\nu}=\frac{x^4 e^x}{(e^x -1)^2},
\EE
where $x$ $=$ $h\nu/k_{\rm B}T_{\mathrm{CMB}}$ $\backsimeq$ $\nu/(56.78\,\mathrm{GHz})$ for $T_{\mathrm{CMB}}=2.725$~K.

To determine the distribution of contamination $s$ we iterate this procedure $10^6$ times for a given cluster mass and redshift and obtain the fraction of clusters above a given value of $s$, as plotted in Fig.~\ref{fig:Contamination1}.  The  color plot in the left panel shows the fraction of clusters with $s \geq 0.1$ for different cluster masses and redshifts.  One can see the contamination -- at any fixed redshift -- is highest for low mass clusters.  This is simply because we have modeled our LF as $M_{500}^{-1}$, implying that the total flux of expected radio galaxies $S_\mathrm{A}$ will scale linearly with the mass of the cluster.  On the other hand, the SZE signature scales as $S_\mathrm{SZE}\propto M_{500}^{5/3}$.  Thus the contamination $s$ in equation~(\ref{eqn:s_contamination}) scales approximately as $s\propto M_{500}^{-2/3}$.  Also, note that at a fixed mass $M_{500}$, the impact of the radio galaxies is highest at low redshift.  This follows because in our preferred model the LF does not evolve with redshift (see discussion of the impact of evolution in section~\ref{sec:redshift_evolution} below), and the SZE flux is approximately constant with redshift \citep[see discussion in][]{majumdar03} while the flux of a source of given luminosity falls as $d_\mathrm{L}^{-2}(z)$ where $d_\mathrm{L}$ is the luminosity distance.  

The plot on the right in Fig.~\ref{fig:Contamination1} shows the fraction of clusters contaminated above a given value of $s$ for the two frequencies relevant for SZE selection at a specific mass and redshift ($M_{500}=3\times10^{14}\rm M_\odot$ and $z=0.25$, which correspond approximately to the lower mass and redshift limits for the SZ-SPT survey).  The contamination is higher at 95~GHz due to the smaller SZE flux at 95~GHz as compared to 150~GHz and the typically higher radio galaxy luminosity at 95~GHz than at 150~GHz.

As the degree of contamination $s$ reaches unity, the cluster -- if unresolved in the SZE maps -- exhibits no net SZE signature.  We calculate that at redshift $z=0.25$ a fraction 0.5 (1.4)~percent of clusters with mass $M_{500}=3\times10^{14}\rm M_{\odot}$ have no net SZE signature in observations at 150 (95)~GHz.   

\citet{lin07} found a much larger contamination by a factor of $\sim6$ at mass $M_{200}=2\times10^{14}\rm M_\odot$ and $z=0.6$.  This discrepancy is rooted in the fact that, as they emphasized, their results involve an extrapolation of the 1.4~GHz LF to 150~GHz using the distribution of spectral indices measured between 1.4~GHz and 4.85~GHz together with an additional break of 0.5 in $\alpha$ at 100~GHz.  With this approach, the rather small fraction of radio galaxies with positive $\alpha$ end up populating a high luminosity portion ($P>10^{26}$~W~Hz$^{-1}$) of the LF that we do not observe in our high frequency sample.

In a more recent study of 139 cluster radio sources selected at 1.4~GHz and observed at 4.9, 8.5, 22 and 43~GHz with the Very Large Array (VLA) \citep{lin09}, the LFs are extrapolated by using spectral indices extracted from 1.4 $\rightarrow$ 4.9 $\rightarrow$ 8.5 $\rightarrow$ 22 $\rightarrow$ 43 $\rightarrow$ 145~GHz. The 145~GHz LF is consistent with our 150~GHz LF within the 1-$\sigma$ model uncertainties. In addition, their estimates for the fraction of missing clusters are similar to our own. \citet{sehgal10} analyzed a full-sky, half-arcminute resolution simulations of the microwave sky matched to the observations from ACT to study the correlation of radio galaxies with SZE clusters.  Their study suggests that at 148~GHz (90~GHz), for clusters with $M_{200} > 10^{14} \rm M_{\odot}$, less than 3 (4) per cent of the clusters have their SZE decrements biased by 20 per cent or more.

\subsection{Incompleteness of SPT-Like Cluster Sample}

To estimate the scale of the effect of cluster radio galaxy contamination on the cluster sample detected in the 2500~deg$^2$ SPT-SZ survey, we first construct the distributions of fractional contamination $s$ at different cluster masses and redshifts. Then we sample from the halo mass function \citep{tinker08,eisenstein98} and use the mass--observable relations to predict the SZE observable with and without the radio galaxy flux biases.  The SZE observable is the detection significance $\xi$, which is related to the halo mass through a two step process.  First, $\xi$ is biased through the multi-scale matched-filter extraction \citep{melin06}, specifically through the selection of the maximum value as a function of position and scale. Thus, it is related to an  unbiased SZE significance $\zeta$ \citep{vanderlinde10}, which is the signal-to-noise at the true, underlying cluster position and the filter scale. The relation between $\xi$ and $\zeta$ is
\BE
 \label{eqn:xi_zeta}
\zeta = \sqrt{\langle\xi\rangle^2-3}.
\EE
Second, the unbiased significance $\zeta$ is related to mass $M_{500}$ as
\BE
\label{eq:zeta_mass}
\zeta = A_{\textrm{SZ}}\left(\frac{M_{500}}{3\times 10^{14}{\rm M_{\odot}}h^{-1}}\right)^{B_{\textrm{SZ}}}\left(\frac{E(z)}{E(0.6)}\right)^{C_{\textrm{SZ}}},
\EE
where $A_{\textrm{SZ}}$ is the normalization, $B_{\textrm{SZ}}$ is the mass power law index, $C_{\textrm{SZ}}$ is the redshift evolution parameter and $E(z)\equiv H(z)/H_0$.  For our calculation we adopt the published values for these parameters \citep{bleem15}.  The fractional intrinsic scatter in the $\zeta -$mass relation, which is assumed to be log-normal and constant as a function of mass and redshift is given as $D_{\textrm{SZ}}\sim0.22$.  Rather than modeling the individual subfields within the SPT-SZ survey, we use a single field with a mean depth scale factor of 1.13 for $A_{\textrm{SZ}}$ \citep[see section 2.1 in][for details about SPT-SZ subfields]{bleem15}.

To select clusters from the mass function, we first integrate the halo mass function over a mass range $10^{14}{\rm M_{\odot}}\le M_{500}\le10^{16}{\rm M_{\odot}}$ and redshift in bins of $\Delta z=$ 0.1 in a redshift range of 0.25 to 1.55 to obtain the expected number of clusters $\langle N_{\rm C}(z_i) \rangle$ in  each redshift bin $z_i$.  We then Poisson sample the number of clusters $N_\mathrm{C}$ in each bin, and for each of these we assign the mass by sampling the mass function.  Given the mass and redshift, we use the $\zeta-$mass relation as in equation~(\ref{eq:zeta_mass}) and the log-normal scatter to calculate the $\zeta$ for each cluster.  We then transform from $\zeta$ to $\xi$ using equation~(\ref{eqn:xi_zeta}) and a normal distribution with standard deviation of unity, which represents the observational noise on the quantity $\xi$.  In the end we apply a $\xi$ based selection exactly as it is done within the real SPT-SZ analysis;  we examine here the threshold $\xi\geq 4.5$.

To study the effect of cluster radio galaxies on the cluster number counts, we adopt the same procedure but introduce a random contaminating flux appropriate for the cluster mass and redshift.  Specifically, we derive the contaminated SZE significance $\zeta_{\rm c}$ as
\BE
\label{eqn:zeta_obs}
\zeta_{\mathrm c} = \zeta (1-s_{\mathrm r}),
\EE
where $s_{\rm r}$ is a randomly selected value of the radio galaxy contamination $s$ drawn from the calculated distribution of $s$ for the given cluster mass and redshift.  Here the $s$ distribution not only accounts for the cluster to cluster variation, but also takes into account the uncertainties in the best fit LF parameters. We then calculate $\xi$ from $\zeta_{\rm c}$ as described above.  After applying the same selection threshold $\xi\geq4.5$, we find that there is a $1.8\pm0.7$~percent reduction in the number of galaxy clusters over the redshift range $0.25\le z\le1.55$ in a 2500~deg$^2$ SPT-like SZE survey.  The error bars are evaluated by generating 100 realizations of the survey and sampling the $s$ distributions as previously described.  The decrease in the number counts as a function of redshift is shown in Fig.~\ref{fig:cluster_counts} for one of the realizations.  The ratio of the recovered number of clusters after contamination $N_{\rm Obs}$ to the number expected without contamination $N$ for these $\xi>4.5$ samples varies from $\sim0.96$ at $z=0.3$ to $\sim0.99$ at $z=1.5$.  Given the size of the current SPT sample \citep{bleem15} the scale of this systematic is small compared to the Poisson sampling noise and therefore not important for recent cosmological studies \citep[e.g.][]{bocquet15, dehaan16}. 

Note that the level of incompleteness presented here is only due to the radio AGN in clusters. In principle, dusty galaxies could also affect the SZE signal. However, we expect the contamination due to dusty galaxies to be minor for clusters in the mass range probed by SPT-SZ, because the galaxy populations are dominated by red sequence galaxies \citep{hennig16}, and in general the number of dusty galaxies identified at the 95 and 150~GHz frequencies within SPT-SZ data is smaller by a factor of $\sim$4 \citep{mocanu13}.  In fact, the majority of these dusty sources are lensed background sources \citep[see][and references there in]{vieira10, mocanu13} because dusty star forming galaxies are very rare within cluster populations.
%and, therefore, the resulting suppression of cluster counts is straightforward to estimate.

There are two other comments of note.  First, gravitational lensing of sources behind the cluster increases their observed flux, making it more likely that they appear in a flux limited sample.  We do not expect this to have any measurable impact on our measured LFs \citep[see, e.g.,][]{chiu16}, and therefore we do not apply any correction.  Second, the mock SPT-SZ survey described here is modeled with cluster selection at 150~GHz only, which differs from the real SPT-SZ survey where information is incorporated from both 95 and 150~GHz. Thus, we expect the contamination in the real SPT-SZ survey to be slightly higher than (but within the error bars) the results presented here.

\subsection{Impact on $\zeta-M_{500}$ Scaling Relation}

\subsubsection{Scaling Relation Parameters}
We examine the bias in the parameters of the $\zeta-$mass relation as described in equation~(\ref{eq:zeta_mass}), caused by AGN contamination in clusters. For this purpose, we take all clusters with $\xi\geq$ 4.5 in a redshift range of 0.25 to 1.55. Using an MCMC and assuming the fixed cosmology used throughout this work, we fit the scaling relation for the $\zeta$ and $\zeta_{\rm c}$ distributions of these clusters to get the best fit parameter values and uncertainties. We find that the shift in the best fit parameters obtained from the uncontaminated signal ($\zeta$) and the AGN contaminated signal ($\zeta_{\rm c}$) is small and is well within the 1-$\sigma$ statistical parameter uncertainties in the two cases.  These shifts are of the order of 1\%, 2\% and 12\% for $A_{\rm SZ}$, $B_{\rm SZ}$ and $C_{\rm SZ}$, respectively. 

In the cosmological analyses presented by the SPT collaboration, we do not assume a perfect knowledge of the $\zeta$-mass scaling relations, but vary these parameters using Gaussian priors. Therefore, a particular bias on the SZE signal caused by radio sources is only important if that bias is large compared to the width of our priors on the scaling relation parameters. The bias we evaluate here is much smaller than the priors we assume in our most recent cosmological analysis \citep[][]{dehaan16}.

\subsubsection{Scatter}

We also examine the contribution of the cluster radio galaxies to the intrinsic scatter in the $\zeta-$mass relation. We calculate the scatter $\sigma_{{\ln}s}$ in the $\zeta_{\rm c}/ \zeta$ (note that $\zeta_{\rm c}/ \zeta\simeq1-s_{\rm r}$) distribution for clusters with $\xi\geq$ 4.5 at different redshifts. The combined distribution in a redshift range of 0.25 to 1.55 has $\sigma_{{\ln}s}\sim 0.028\pm0.004$.  As noted previously, the calibrated total intrinsic scatter is 22~percent in the $\zeta-$mass relation, and therefore this contribution from cluster radio galaxy contamination plays no significant role in explaining the total observed scatter in the mass--observable relation we employ for the SPT-SZ sample.

\subsection{Redshift Evolution of the Luminosity Function}
\label{sec:redshift_evolution}
So far in our analysis for contamination, we have assumed that the LF at 150 $\rm GHz$ does not evolve with redshift.  Our analysis (see right panel of Fig.~\ref{fig:L_function}) supports this assumption at lower observing frequency.  The limited sample of galaxy clusters we have in the SPT region with a low median redshift of 0.1 makes it difficult to probe for evolution to higher redshift. This non-evolution, however, is very well supported by a number of previous studies. For example, \citet{stocke99} compared the 1.4~GHz observations of 19 X-ray selected galaxy clusters in the redshift range of 0.3 to 0.8, with nearby clusters from \citet{ledlow96} and found no evidence of evolution. Similar results were obtained by \citet{branchesi06} for a sample of 18 X-ray selected galaxy clusters in the same redshift range. \citet{gralla11} constrain the evolution of the bright central radio source population in galaxy clusters from redshift 0.35 to 0.95 by statistically matching FIRST radio sources \citep{becker95} with 618 galaxy clusters from a uniformly, optically selected sample RCS1 \citep{gladders05} and find 0.14$\pm$0.02 and 0.10$\pm$0.02 radio sources per cluster in the range of 0.35$<$$z$$<$0.65 and 0.65$<$$z$$<$0.95, respectively.  \citet{fassbender11} study a sample of 22 clusters at 0.9$<$$z$$<$1.6 and show that 30 per cent of them have a central 1.4~GHz radio source.  Given the small sample used in this study, the results are consistent with those from the more comprehensive studies already mentioned. 
%Comparing to the fraction in low redshift clusters \citep[studied in][]{lin07,best07,mittal09}, they suggest a factor of 2.5 to 5 increase in the fraction of clusters containing luminous radio galaxies at high redshifts. However, they conclude that this rise in the number of radio-loud cluster fraction with redshift over the probed interval is limited by statistics given their small sample.  
%Thus, there are no clear indications that the low frequency radio LF evolves strongly with redshift.

%
\begin{figure}
\vskip-0.2in
\centering
\includegraphics[width=9cm, height=8cm]{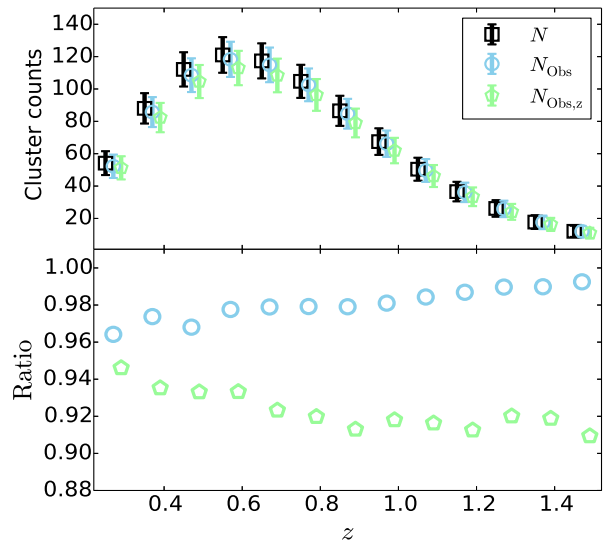}
\vskip-0.1in
\caption{The estimated decrease in the observed number of galaxy clusters ($N_{\rm Obs}$) due to the point source contamination as compared to the theoretical number counts ($N$) for 2500 $\rm deg^2$ of the SPT survey with $\xi_{\rm r}\geq$ 4.5. The decrease in the observed number of clusters ($N_{\rm Obs,z}$) is also shown for a possible redshift evolution of the form $(1+z)^{2.5}$ in the number of point sources.}
\label{fig:cluster_counts}
\end{figure}

In a recent study, \citet{pracy16}  derived 1.4~GHz LFs for radio AGN separated into Low Excitation Radio Galaxies (LERGs) and High Excitation Radio Galaxies (HERGs), in the three redshift bins 0.005 < z < 0.3, 0.3 < z < 0.5 and 0.5 < z < 0.75.  They found that the LERG population displays little or no evolution $(1+z)^{0.06^{+0.16}_{-0.18}}$ over this redshift range, while the HERG population evolves more rapidly as $(1+z)^{2.93^{+0.46}_{-0.47}}$ (assuming pure density evolution in both cases).  HERGs have bluer color and lower 4000 \AA{} breaks, which are indications of ongoing star formation activity.
% as is also seen in similar radio quiet AGN \citep{kauffmann03a}. 
LERGs, however, appear to be preferentially located at the centers of groups or clusters and are fueled by accretion from their hot gas haloes \citep{kauffmann08, lin10, best12}.  Thus, the LFs presented in our work are presumably dominated by the LERG population and would therefore not be expected to evolve strongly with redshift.

Directly constraining the redshift evolution of the cluster radio galaxy LF at high frequency will require a larger sample of clusters extending to high redshift.  In the current analysis we simply bracket the range of possible redshift evolution by examining incompleteness in the case where the radio galaxy number density increases with redshift as $\phi\propto(1+z)^{2.5}$.  We find that in this case there are $5.6\pm1$~percent of the clusters in the 2500~deg$^2$ of the SPT survey at 150~GHz that would be expected to fall out of the $\xi>4.5$ selected sample.  Our estimated change in the number of galaxy clusters with this kind of extreme evolution model is also shown in Fig.~\ref{fig:cluster_counts} in different redshift bins, along with the change we calculate with no redshift evolution.

In this evolutionary scenario we find that the shift in the best fit parameters of the $\zeta-$mass relation, obtained by comparing uncontaminated signal ($\zeta$) and contaminated signal ($\zeta_{\rm c}$) is within the 1-$\sigma$ parameter constraints for $B_{\rm SZ}$ and $C_{\rm SZ}$. The best fit value of $A_{\rm SZ}$ is however biased low by 3 percent in the AGN contaminated case.  We also calculate that $4.8\pm0.5$~percent of the scatter in the $\zeta-$mass relation would come from the cluster to cluster variation in contamination due to cluster radio galaxies.  This is still small compared to the empirically constrained scatter of 22~percent.

\section{Systematics}
\label{sec:systematics}
In this section, we discuss the impact of systematic uncertainties in the cluster masses and the radio galaxy fluxes on our results.

\subsection{MCXC Cluster Mass Uncertainties}
\label{sec:MCXC_masses}
Because MCXC is a heterogeneous catalog compiled from different ROSAT X-ray Sky Survey-based catalogs (see section~\ref{sec:MCXC}), we do not have the information about the mass uncertainties of these clusters. However it is known that the X-ray luminosity--mass relation exhibits a scatter on the order of 40~percent \citep{vikhlinin09b, mantz10b}.   Moreover, there is ongoing discussion in the literature about the difference between X-ray hydrostatic masses and other mass estimates, including velocity dispersions, weak gravitational lensing and calibration through the cluster mass function \citep[see][]{bocquet15}.  Thus we construct our LFs by increasing and decreasing the masses of all clusters by 50~percent to see the maximum impact on our results. The 50~percent decrease in cluster masses affects the LF in three ways: (1) there are fewer sources inside the clusters because the virial radius is smaller, (2) the SZE flux correction for the point sources inside clusters is smaller by $\approx50$~percent and  (3) the normalization of the LF rises as we scale the observed population as $M_{500}^{-1}$. We find that the best fit LFs produced by increasing and decreasing cluster masses are within the 2$\sigma$ model uncertainties of the best fit LF constructed from the central estimates of the cluster masses.  Accordingly, LFs from a 50~percent increase and decrease in the masses are found to lead to incompleteness in an SPT-like sample of  $1.4\pm0.8$~percent and $3.0\pm0.7$~percent, respectively, in a redshift range of 0.25 to 1.55.  This compares to the $1.8\pm0.7$~percent incompleteness using the published MCXC masses.

\subsection{Radio Galaxy Flux Uncertainties}
\label{sec:flux_uncertainties}
To fit the luminosity function we compare the observations to the model to determine the likelihood without taking into account the uncertainties in the point source fluxes. To account for the flux uncertainties we compare the observations to the LF model after convolving it with the appropriate flux uncertainties.   Because each source has a different flux uncertainty (corresponding to the uncertainty in the luminosity) this is done by taking a Gaussian weighted average over the relevant part of the model (i.e we extracted a convolved value of the model) only for the luminosity bins where there are measured galaxies.  For the empty bins we convolve the model with the luminosity uncertainty equivalent to the maximum uncertainty from any source.  We see a small difference in the best fit parameters for the LF, which is consistent within the 1-$\sigma$ parameter uncertainties from the LF of the unconvolved model.

In addition, the radio galaxy source counts are a steep function of flux and the uncertainties in flux could potentially lead to some bias in the LF measurements. Following \cite{mortonson11}, we estimate this bias at the flux cut to be much smaller than the statistical errors on the LFs; thus this effect has no impact on the contamination estimates.

\section{Conclusions}
\label{sec:conclusions}

We use the MCXC catalog of galaxy clusters, the SUMSS catalog of radio galaxies, and the SPT-SZ survey maps to measure the overdensity of radio galaxies associated with clusters.  We construct radio galaxy LFs and radial profiles at 843~MHz, 95~GHz, 150~GHz and 220~GHz.  The MCXC systems in the SPT-SZ and SUMSS regions have a median redshift $z\sim0.1$, and the highest redshift system is at $z=0.686$.  There are 139 MCXC objects in the SPT-SZ region and 333 in the SUMSS region; they span the mass range from groups to clusters with a median mass $M_{500}=1.5\times10^{14}\rm{M_\odot}$ and $M_{500}=1.7\times10^{14}\rm{M_\odot}$ in the SPT-SZ and SUMSS regions, respectively.

To construct LFs at high frequencies, we examine SPT maps at the locations of SUMSS sources, extracting the high frequency fluxes and correcting for the cluster SZE flux at 95 and 150~GHz.  We compare this sample with the 150~GHz sample with uncorrected fluxes to examine the impact of SZE flux biases, showing that they are significant -- especially for high redshift clusters that are more compact on the sky and for higher mass clusters that have stronger SZE signatures.  In essence, it is more challenging to find cluster radio galaxies at high frequency in high redshift and high mass clusters, because the SZE signature is biasing their fluxes low. 

We use the SUMSS selected sources with fluxes measured at SPT frequencies and correct for SZE flux bias (at 95 and 150~GHz) to construct the cluster radio galaxy sample for further analysis.  We find that the radial profile is centrally concentrated, consistent with an NFW model with concentration $c=108^{+107}_{-48}$.  We examine the spectral indices of the radio galaxy population, finding that the spectral index $\alpha$ measured between 95 and 150~GHz is steeper than that measured between 843~MHz and these high frequencies.  We construct the LFs and find best fit parametrizations within the context of \citet{condon02} models. In doing so, we assume the overdensity of radio galaxies toward a cluster is at the redshift of the cluster, and we apply a $k$-correction using the spectral indices extracted from the sample.  Above a luminosity of $10^{21}$~W~Hz$^{-1}$ the 150~GHz LF has roughly half the amplitude of the 95~GHz LF (see Table~\ref{tab:LFparameters} and Fig.~\ref{fig:LF_different_neu}).  The amplitude of the 843~MHz LF is approximately one order of magnitude higher than the amplitude of the high frequency LFs.  Our high frequency radio galaxy sample is not large enough to constrain redshift or mass trends in the radio galaxy LF.

We use the measured high frequency cluster radio galaxy LFs to examine the effect of the contaminating flux on the SZE signatures of galaxy clusters. To do that, we use the LF for a given cluster mass and redshift to obtain the number and flux of cluster radio galaxies, sampling $10^6$ times to recover the full range of behavior of the cluster radio galaxies within the clusters.  We define a quantity called the contamination $s$, which is the absolute value of the ratio of the total cluster radio galaxy flux from all the radio galaxies with power $>10^{21}$~W~Hz$^{-1}$ to the total SZE flux of that cluster within $r_{200}$.  With this information we calculate the fraction of clusters with $s\simeq 1$, where the total cluster radio galaxy flux in a cluster is equivalent to the negative SZE flux.  We find that 0.5 and 1.4~percent of clusters meet this criterion for cluster mass $M_{500}=3\times10^{14}\rm M_{\odot}$ and redshift $z=0.25$ at 150 and 95~GHz, respectively.

To estimate the impact of cluster radio galaxies on the cluster sample from the SPT-SZ 2500~deg$^2$ survey at 150~GHz, we use the theoretically predicted mass function to produce 100 mock cluster samples.  We then compare the $\xi>4.5$ cluster samples with and without cluster radio galaxies.  We find that around $1.8\pm0.7$~percent of clusters would be lost from the sample in a redshift range of 0.25 to 1.55 in the 2500~deg$^2$ SPT-SZ survey. 

We evaluate the bias in the parameters of the $\zeta-$mass relation caused by radio galaxy contamination and find a small shift in the mean parameter values which is well within the current 1-$\sigma$ parameter constraints.  We also calculate the contribution of the cluster radio galaxy contamination to the intrinsic scatter in the $\zeta-$mass relation for the observed clusters, finding that cluster radio galaxies contribute a scatter of $2.8\pm0.4$~percent out of a total empirically calibrated $\sim$22~percent scatter.

Finally, we note that with the MCXC sample we cannot place strong constraints on the redshift evolution of the high frequency radio galaxy LF.  We review previous findings at 1.4~GHz, none of which provide evidence for strong redshift evolution of the cluster radio galaxy LF.  We attempt to bracket the impact of possible redshift evolution by adopting a radio galaxy LF evolution in the number of point sources of the form $(1+z)^{2.5}$, showing that at 150~GHz there could be a $5.6\pm1$~percent incompleteness in a $\xi>4.5$ SPT-SZ like SZE selected cluster sample.

It has been noted that in the SPT and Planck SZE selected cluster samples there is a preference for higher cluster masses when these masses are calibrated in conjunction with external cosmological constraints \citep[e.g.][]{planck15cosm} in comparison to direct calibration using weak lensing, velocity dispersions, CMB lensing or X-ray hydrostatic masses \citep[see fig.~2 and fig.~8, respectively, in][]{bocquet15,planck15}.  Incompleteness in the SZE selected cluster samples is one of several possible effects, including systematic mass biases or even biases in the adopted theoretical mass function \citep[see][]{bocquet16} that could contribute to this preference.  Given the results of our high frequency cluster radio galaxy study, it appears that incompleteness in SZE selected cluster samples due to radio AGN is too small to be playing an important role.  
%Further study of this phenomena is warranted. 

Clearly, a larger sample of non-SZE selected clusters with accurate mass estimates and spanning a larger redshift range is needed to resolve the issues of redshift evolution of the radio galaxy LF and to improve the constraints on the LFs at 150 and 95~GHz.  More precise measurements of high frequency radio galaxy LFs will also help us to accurately estimate the incompleteness in the ongoing or upcoming SZE surveys like SPTpol \citep{bleem12a}, SPT-3G \citep{benson14} and CMB-S4 \citep{abazajian15}, which are all expected to be sensitive to lower mass clusters.  We are exploring such samples using the Dark Energy Survey \citep{DES05} today and are looking forward to the opportunity to examine this population of galaxies in the upcoming eROSITA X-ray survey \citep{merloni12,predehl14}.  

\section*{Acknowledgements}

We acknowledge the support of the International Max Planck Research School on Astrophysics of the Ludwig-Maximilians-Universit\"at, the Max-Planck-Gesellschaft Faculty Fellowship program at the Max Planck Institute for Extraterrestrial Physics, the DFG Cluster of Excellence ``Origin and Structure of the Universe'', the Transregio program TR33 ``The Dark Universe'' and the Ludwig-Maximilians-Universit\"at. The data processing has been carried out on the computing facilities of the Computational Center for Particle and Astrophysics (C2PAP), located at the Leibniz Supercomputer Center (LRZ).  

The South Pole Telescope is supported by the National Science Foundation through grant PLR-1248097. Partial support is also provided by the NSF Physics Frontier Center grant PHY-1125897 to the Kavli Institute of Cosmological Physics at the University of Chicago, the Kavli Foundation and the Gordon and Betty Moore Foundation grant GBMF 947.  
B. A. Benson is supported by the Fermi Research Alliance, LLC under Contract No. De-AC02-07CH11359 with the United States Department of Energy.

\bibliographystyle{mnras}
\bibliography{LF_SPT}

\begin{thebibliography}{}
\makeatletter
\relax
\def\mn@urlcharsother{\let\do\@makeother \do\$\do\&\do\#\do\^\do\_\do\%\do\~}
\def\mn@doi{\begingroup\mn@urlcharsother \@ifnextchar [ {\mn@doi@}
  {\mn@doi@[]}}
\def\mn@doi@[#1]#2{\def\@tempa{#1}\ifx\@tempa\@empty \href
  {http://dx.doi.org/#2} {doi:#2}\else \href {http://dx.doi.org/#2} {#1}\fi
  \endgroup}
\def\mn@eprint#1#2{\mn@eprint@#1:#2::\@nil}
\def\mn@eprint@arXiv#1{\href {http://arxiv.org/abs/#1} {{\tt arXiv:#1}}}
\def\mn@eprint@dblp#1{\href {http://dblp.uni-trier.de/rec/bibtex/#1.xml}
  {dblp:#1}}
\def\mn@eprint@#1:#2:#3:#4\@nil{\def\@tempa {#1}\def\@tempb {#2}\def\@tempc
  {#3}\ifx \@tempc \@empty \let \@tempc \@tempb \let \@tempb \@tempa \fi \ifx
  \@tempb \@empty \def\@tempb {arXiv}\fi \@ifundefined
  {mn@eprint@\@tempb}{\@tempb:\@tempc}{\expandafter \expandafter \csname
  mn@eprint@\@tempb\endcsname \expandafter{\@tempc}}}

\bibitem[\protect\citeauthoryear{{Abazajian} et~al.,}{{Abazajian}
  et~al.}{2015}]{abazajian15}
{Abazajian} K.~N.,  et~al., 2015, \mn@doi [Astroparticle Physics]
  {10.1016/j.astropartphys.2014.05.014}, \href
  {http://adsabs.harvard.edu/abs/2015APh....63...66A} {63, 66}

\bibitem[\protect\citeauthoryear{{Arnaud}, {Pratt}, {Piffaretti},
  {B{\"o}hringer}, {Croston}  \& {Pointecouteau}}{{Arnaud}
  et~al.}{2010}]{arnaud10}
{Arnaud} M.,  {Pratt} G.~W.,  {Piffaretti} R.,  {B{\"o}hringer} H.,  {Croston}
  J.~H.,   {Pointecouteau} E.,  2010, \mn@doi [\aap]
  {10.1051/0004-6361/200913416}, \href
  {http://adsabs.harvard.edu/abs/2010A%26A...517A..92A} {517, A92+}

\bibitem[\protect\citeauthoryear{{Bartelmann}}{{Bartelmann}}{1996}]{bartelmann96}
{Bartelmann} M.,  1996, \aap, \href
  {http://adsabs.harvard.edu/abs/1996A%26A...313..697B} {313, 697}

\bibitem[\protect\citeauthoryear{{Becker}, {White}  \& {Helfand}}{{Becker}
  et~al.}{1995}]{becker95}
{Becker} R.~H.,  {White} R.~L.,   {Helfand} D.~J.,  1995, \mn@doi [\apj]
  {10.1086/176166}, \href {http://adsabs.harvard.edu/abs/1995ApJ...450..559B}
  {450, 559}

\bibitem[\protect\citeauthoryear{{Benson} et~al.,}{{Benson}
  et~al.}{2013}]{benson13}
{Benson} B.~A.,  et~al., 2013, \mn@doi [\apj] {10.1088/0004-637X/763/2/147},
  \href {http://adsabs.harvard.edu/abs/2013ApJ...763..147B} {763, 147}

\bibitem[\protect\citeauthoryear{{Benson} et~al.,}{{Benson}
  et~al.}{2014}]{benson14}
{Benson} B.~A.,  et~al., 2014, in Millimeter, Submillimeter, and Far-Infrared
  Detectors and Instrumentation for Astronomy VII. p. 91531P (\mn@eprint
  {arXiv} {1407.2973}), \mn@doi{10.1117/12.2057305}

\bibitem[\protect\citeauthoryear{{Best} \& {Heckman}}{{Best} \&
  {Heckman}}{2012}]{best12}
{Best} P.~N.,  {Heckman} T.~M.,  2012, \mn@doi [\mnras]
  {10.1111/j.1365-2966.2012.20414.x}, \href
  {http://adsabs.harvard.edu/abs/2012MNRAS.421.1569B} {421, 1569}

\bibitem[\protect\citeauthoryear{{Bleem} et~al.,}{{Bleem}
  et~al.}{2012}]{bleem12a}
{Bleem} L.,  et~al., 2012, \mn@doi [Journal of Low Temperature Physics]
  {10.1007/s10909-012-0505-y}, \href
  {http://adsabs.harvard.edu/abs/2012JLTP..tmp..196B} {p.~196}

\bibitem[\protect\citeauthoryear{{Bleem} et~al.,}{{Bleem}
  et~al.}{2015}]{bleem15}
{Bleem} L.~E.,  et~al., 2015, \mn@doi [\apjs] {10.1088/0067-0049/216/2/27},
  \href {http://adsabs.harvard.edu/abs/2015ApJS..216...27B} {216, 27}

\bibitem[\protect\citeauthoryear{{Bock}, {Large}  \& {Sadler}}{{Bock}
  et~al.}{1999}]{bock99}
{Bock} D.~C.-J.,  {Large} M.~I.,   {Sadler} E.~M.,  1999, \mn@doi [\aj]
  {10.1086/300786}, \href {http://adsabs.harvard.edu/abs/1999AJ....117.1578B}
  {117, 1578}

\bibitem[\protect\citeauthoryear{{Bocquet} et~al.,}{{Bocquet}
  et~al.}{2015}]{bocquet15}
{Bocquet} S.,  et~al., 2015, \mn@doi [\apj] {10.1088/0004-637X/799/2/214},
  \href {http://adsabs.harvard.edu/abs/2015ApJ...799..214B} {799, 214}

\bibitem[\protect\citeauthoryear{{Bocquet}, {Saro}, {Dolag}  \&
  {Mohr}}{{Bocquet} et~al.}{2016}]{bocquet16}
{Bocquet} S.,  {Saro} A.,  {Dolag} K.,   {Mohr} J.~J.,  2016, \mn@doi [\mnras]
  {10.1093/mnras/stv2657}, \href
  {http://adsabs.harvard.edu/abs/2016MNRAS.456.2361B} {456, 2361}

\bibitem[\protect\citeauthoryear{{B{\"o}hringer} et~al.,}{{B{\"o}hringer}
  et~al.}{2000}]{bohringer00}
{B{\"o}hringer} H.,  et~al., 2000, \mn@doi [\apjs] {10.1086/313427}, \href
  {http://adsabs.harvard.edu/abs/2000ApJS..129..435B} {129, 435}

\bibitem[\protect\citeauthoryear{{B{\"o}hringer} et~al.,}{{B{\"o}hringer}
  et~al.}{2004}]{bohringer04}
{B{\"o}hringer} H.,  et~al., 2004, \mn@doi [\aap] {10.1051/0004-6361:20034484},
  \href {http://adsabs.harvard.edu/abs/2004A%26A...425..367B} {425, 367}

\bibitem[\protect\citeauthoryear{{Branchesi}, {Gioia}, {Fanti}, {Fanti}  \&
  {Perley}}{{Branchesi} et~al.}{2006}]{branchesi06}
{Branchesi} M.,  {Gioia} I.~M.,  {Fanti} C.,  {Fanti} R.,   {Perley} R.,  2006,
  \mn@doi [\aap] {10.1051/0004-6361:20053767}, \href
  {http://adsabs.harvard.edu/abs/2006A%26A...446...97B} {446, 97}

\bibitem[\protect\citeauthoryear{{Burenin}, {Vikhlinin}, {Hornstrup},
  {Ebeling}, {Quintana}  \& {Mescheryakov}}{{Burenin} et~al.}{2007}]{burenin07}
{Burenin} R.~A.,  {Vikhlinin} A.,  {Hornstrup} A.,  {Ebeling} H.,  {Quintana}
  H.,   {Mescheryakov} A.,  2007, \mn@doi [\apjs] {10.1086/519457}, \href
  {http://adsabs.harvard.edu/abs/2007ApJS..172..561B} {172, 561}

\bibitem[\protect\citeauthoryear{{Carlstrom} et~al.,}{{Carlstrom}
  et~al.}{2011}]{carlstrom11}
{Carlstrom} J.~E.,  et~al., 2011, \mn@doi [\pasp] {10.1086/659879}, \href
  {http://adsabs.harvard.edu/abs/2011PASP..123..568C} {123, 568}

\bibitem[\protect\citeauthoryear{{Cash}}{{Cash}}{1979}]{cash79}
{Cash} W.,  1979, \mn@doi [\apj] {10.1086/156922}, \href
  {http://adsabs.harvard.edu/abs/1979ApJ...228..939C} {228, 939}

\bibitem[\protect\citeauthoryear{{Chiu} et~al.,}{{Chiu} et~al.}{2016}]{chiu16}
{Chiu} I.,  et~al., 2016, \mn@doi [\mnras] {10.1093/mnras/stw190}, \href
  {http://adsabs.harvard.edu/abs/2016MNRAS.457.3050C} {457, 3050}

\bibitem[\protect\citeauthoryear{{Coble} et~al.,}{{Coble}
  et~al.}{2007}]{coble07}
{Coble} K.,  et~al., 2007, \mn@doi [\aj] {10.1086/519973}, \href
  {http://adsabs.harvard.edu/abs/2007AJ....134..897C} {134, 897}

\bibitem[\protect\citeauthoryear{{Condon}}{{Condon}}{1992}]{condon92}
{Condon} J.~J.,  1992, \mn@doi [\araa] {10.1146/annurev.aa.30.090192.003043},
  \href {http://adsabs.harvard.edu/abs/1992ARA%26A..30..575C} {30, 575}

\bibitem[\protect\citeauthoryear{{Condon}, {Cotton}, {Greisen}, {Yin},
  {Perley}, {Taylor}  \& {Broderick}}{{Condon} et~al.}{1998}]{condon98}
{Condon} J.~J.,  {Cotton} W.~D.,  {Greisen} E.~W.,  {Yin} Q.~F.,  {Perley}
  R.~A.,  {Taylor} G.~B.,   {Broderick} J.~J.,  1998, \mn@doi [\aj]
  {10.1086/300337}, \href {http://adsabs.harvard.edu/abs/1998AJ....115.1693C}
  {115, 1693}

\bibitem[\protect\citeauthoryear{{Condon}, {Cotton}  \& {Broderick}}{{Condon}
  et~al.}{2002}]{condon02}
{Condon} J.~J.,  {Cotton} W.~D.,   {Broderick} J.~J.,  2002, \mn@doi [\aj]
  {10.1086/341650}, \href {http://adsabs.harvard.edu/abs/2002AJ....124..675C}
  {124, 675}

\bibitem[\protect\citeauthoryear{{Cooray}, {Grego}, {Holzapfel}, {Joy}  \&
  {Carlstrom}}{{Cooray} et~al.}{1998}]{cooray98}
{Cooray} A.~R.,  {Grego} L.,  {Holzapfel} W.~L.,  {Joy} M.,   {Carlstrom}
  J.~E.,  1998, \aj, 115, 1388

\bibitem[\protect\citeauthoryear{{Cruddace} et~al.,}{{Cruddace}
  et~al.}{2002}]{cruddace02}
{Cruddace} R.,  et~al., 2002, \mn@doi [\apjs] {10.1086/324519}, \href
  {http://adsabs.harvard.edu/abs/2002ApJS..140..239C} {140, 239}

\bibitem[\protect\citeauthoryear{{DES Collaboration}}{{DES
  Collaboration}}{2005}]{DES05}
{DES Collaboration} 2005, preprint, \href
  {http://adsabs.harvard.edu/abs/2005astro.ph.10346T} {} (\mn@eprint {arXiv}
  {astro-ph/0510346})

\bibitem[\protect\citeauthoryear{{Duffy}, {Schaye}, {Kay}  \& {Dalla
  Vecchia}}{{Duffy} et~al.}{2008}]{duffy08}
{Duffy} A.~R.,  {Schaye} J.,  {Kay} S.~T.,   {Dalla Vecchia} C.,  2008, \mn@doi
  [\mnras] {10.1111/j.1745-3933.2008.00537.x}, \href
  {http://adsabs.harvard.edu/abs/2008MNRAS.390L..64D} {390, L64}

\bibitem[\protect\citeauthoryear{{Ebeling}, {Edge}, {Bohringer}, {Allen},
  {Crawford}, {Fabian}, {Voges}  \& {Huchra}}{{Ebeling}
  et~al.}{1998}]{ebeling98}
{Ebeling} H.,  {Edge} A.~C.,  {Bohringer} H.,  {Allen} S.~W.,  {Crawford}
  C.~S.,  {Fabian} A.~C.,  {Voges} W.,   {Huchra} J.~P.,  1998, \mn@doi
  [\mnras] {10.1046/j.1365-8711.1998.01949.x}, \href
  {http://adsabs.harvard.edu/abs/1998MNRAS.301..881E} {301, 881}

\bibitem[\protect\citeauthoryear{{Ebeling}, {Edge}, {Allen}, {Crawford},
  {Fabian}  \& {Huchra}}{{Ebeling} et~al.}{2000}]{ebeling00}
{Ebeling} H.,  {Edge} A.~C.,  {Allen} S.~W.,  {Crawford} C.~S.,  {Fabian}
  A.~C.,   {Huchra} J.~P.,  2000, \mn@doi [\mnras]
  {10.1046/j.1365-8711.2000.03549.x}, \href
  {http://adsabs.harvard.edu/abs/2000MNRAS.318..333E} {318, 333}

\bibitem[\protect\citeauthoryear{{Ebeling}, {Edge}  \& {Henry}}{{Ebeling}
  et~al.}{2001}]{ebeling01}
{Ebeling} H.,  {Edge} A.~C.,   {Henry} J.~P.,  2001, \mn@doi [\apj]
  {10.1086/320958}, \href {http://adsabs.harvard.edu/abs/2001ApJ...553..668E}
  {553, 668}

\bibitem[\protect\citeauthoryear{{Ebeling}, {Mullis}  \& {Tully}}{{Ebeling}
  et~al.}{2002}]{ebeling02}
{Ebeling} H.,  {Mullis} C.~R.,   {Tully} R.~B.,  2002, \mn@doi [\apj]
  {10.1086/343790}, \href {http://adsabs.harvard.edu/abs/2002ApJ...580..774E}
  {580, 774}

\bibitem[\protect\citeauthoryear{{Eisenstein} \& {Hu}}{{Eisenstein} \&
  {Hu}}{1998}]{eisenstein98}
{Eisenstein} D.~J.,  {Hu} W.,  1998, \mn@doi [\apj] {10.1086/305424}, \href
  {http://adsabs.harvard.edu/abs/1998ApJ...496..605E} {496, 605}

\bibitem[\protect\citeauthoryear{{Everett et al. in}}{{Everett et al.
  in}}{prep}]{everett16}
{Everett et al. in} prep., preprint (\mn@eprint {arXiv} {9999.9999})

\bibitem[\protect\citeauthoryear{{Fassbender} et~al.,}{{Fassbender}
  et~al.}{2011}]{fassbender11}
{Fassbender} R.,  et~al., 2011, \mn@doi [New Journal of Physics]
  {10.1088/1367-2630/13/12/125014}, \href
  {http://adsabs.harvard.edu/abs/2011NJPh...13l5014F} {13, 125014}

\bibitem[\protect\citeauthoryear{{Foreman-Mackey}, {Hogg}, {Lang}  \&
  {Goodman}}{{Foreman-Mackey} et~al.}{2013}]{mackey13}
{Foreman-Mackey} D.,  {Hogg} D.~W.,  {Lang} D.,   {Goodman} J.,  2013, \mn@doi
  [\pasp] {10.1086/670067}, \href
  {http://adsabs.harvard.edu/abs/2013PASP..125..306F} {125, 306}

\bibitem[\protect\citeauthoryear{{Fowler} et~al.,}{{Fowler}
  et~al.}{2007}]{fowler07}
{Fowler} J.~W.,  et~al., 2007, \mn@doi [\ao] {10.1364/AO.46.003444}, \href
  {http://adsabs.harvard.edu/abs/2007ApOpt..46.3444F} {46, 3444}

\bibitem[\protect\citeauthoryear{{Gioia} \& {Luppino}}{{Gioia} \&
  {Luppino}}{1994}]{gioia94}
{Gioia} I.~M.,  {Luppino} G.~A.,  1994, \mn@doi [\apjs] {10.1086/192083}, \href
  {http://adsabs.harvard.edu/abs/1994ApJS...94..583G} {94, 583}

\bibitem[\protect\citeauthoryear{{Gladders} \& {Yee}}{{Gladders} \&
  {Yee}}{2005}]{gladders05}
{Gladders} M.~D.,  {Yee} H.~K.~C.,  2005, \mn@doi [\apjs] {10.1086/427327},
  \href {http://adsabs.harvard.edu/abs/2005ApJS..157....1G} {157, 1}

\bibitem[\protect\citeauthoryear{{Gralla}, {Gladders}, {Yee}  \&
  {Barrientos}}{{Gralla} et~al.}{2011}]{gralla11}
{Gralla} M.~B.,  {Gladders} M.~D.,  {Yee} H.~K.~C.,   {Barrientos} L.~F.,
  2011, \mn@doi [\apj] {10.1088/0004-637X/734/2/103}, \href
  {http://adsabs.harvard.edu/abs/2011ApJ...734..103G} {734, 103}

\bibitem[\protect\citeauthoryear{{Gregory}, {Scott}, {Douglas}  \&
  {Condon}}{{Gregory} et~al.}{1996}]{gregory96}
{Gregory} P.~C.,  {Scott} W.~K.,  {Douglas} K.,   {Condon} J.~J.,  1996,
  \mn@doi [\apjs] {10.1086/192282}, \href
  {http://adsabs.harvard.edu/abs/1996ApJS..103..427G} {103, 427}

\bibitem[\protect\citeauthoryear{{Griffith} \& {Wright}}{{Griffith} \&
  {Wright}}{1993}]{griffith93}
{Griffith} M.~R.,  {Wright} A.~E.,  1993, \aj, 105, 1666

\bibitem[\protect\citeauthoryear{{Hasselfield} et~al.,}{{Hasselfield}
  et~al.}{2013}]{hasselfield13}
{Hasselfield} M.,  et~al., 2013, \mn@doi [\jcap]
  {10.1088/1475-7516/2013/07/008}, \href
  {http://adsabs.harvard.edu/abs/2013JCAP...07..008H} {7, 008}

\bibitem[\protect\citeauthoryear{{Hennig} et~al.,}{{Hennig}
  et~al.}{2016}]{hennig16}
{Hennig} C.,  et~al., 2016, preprint, \href
  {http://adsabs.harvard.edu/abs/2016arXiv160400988H} {} (\mn@eprint {arXiv}
  {1604.00988})

\bibitem[\protect\citeauthoryear{{Henry}}{{Henry}}{2004}]{henry04}
{Henry} J.~P.,  2004, \mn@doi [\apj] {10.1086/421336}, \href
  {http://adsabs.harvard.edu/abs/2004ApJ...609..603H} {609, 603}

\bibitem[\protect\citeauthoryear{{Henry}, {Mullis}, {Voges}, {B{\"o}hringer},
  {Briel}, {Gioia}  \& {Huchra}}{{Henry} et~al.}{2006}]{henry06}
{Henry} J.~P.,  {Mullis} C.~R.,  {Voges} W.,  {B{\"o}hringer} H.,  {Briel}
  U.~G.,  {Gioia} I.~M.,   {Huchra} J.~P.,  2006, \mn@doi [\apjs]
  {10.1086/498749}, \href {http://adsabs.harvard.edu/abs/2006ApJS..162..304H}
  {162, 304}

\bibitem[\protect\citeauthoryear{{H{\"o}gbom}}{{H{\"o}gbom}}{1974}]{hogbom74}
{H{\"o}gbom} J.~A.,  1974, \aaps, 15, 417

\bibitem[\protect\citeauthoryear{{Horner}, {Perlman}, {Ebeling}, {Jones},
  {Scharf}, {Wegner}, {Malkan}  \& {Maughan}}{{Horner}
  et~al.}{2008}]{horner08a}
{Horner} D.~J.,  {Perlman} E.~S.,  {Ebeling} H.,  {Jones} L.~R.,  {Scharf}
  C.~A.,  {Wegner} G.,  {Malkan} M.,   {Maughan} B.,  2008, \mn@doi [\apjs]
  {10.1086/529494}, \href {http://adsabs.harvard.edu/abs/2008ApJS..176..374H}
  {176, 374}

\bibitem[\protect\citeauthoryear{{Ivison} et~al.,}{{Ivison}
  et~al.}{2007}]{ivison07}
{Ivison} R.~J.,  et~al., 2007, \mn@doi [\mnras]
  {10.1111/j.1365-2966.2007.12044.x}, \href
  {http://adsabs.harvard.edu/abs/2007MNRAS.380..199I} {380, 199}

\bibitem[\protect\citeauthoryear{{Kauffmann}, {Heckman}  \& {Best}}{{Kauffmann}
  et~al.}{2008}]{kauffmann08}
{Kauffmann} G.,  {Heckman} T.~M.,   {Best} P.~N.,  2008, \mn@doi [\mnras]
  {10.1111/j.1365-2966.2007.12752.x}, \href
  {http://adsabs.harvard.edu/abs/2008MNRAS.384..953K} {384, 953}

\bibitem[\protect\citeauthoryear{{Kocevski}, {Ebeling}, {Mullis}  \&
  {Tully}}{{Kocevski} et~al.}{2007}]{kocevski07}
{Kocevski} D.~D.,  {Ebeling} H.,  {Mullis} C.~R.,   {Tully} R.~B.,  2007,
  \mn@doi [\apj] {10.1086/513303}, \href
  {http://adsabs.harvard.edu/abs/2007ApJ...662..224K} {662, 224}

\bibitem[\protect\citeauthoryear{{Ledlow} \& {Owen}}{{Ledlow} \&
  {Owen}}{1996}]{ledlow96}
{Ledlow} M.~J.,  {Owen} F.~N.,  1996, \mn@doi [\aj] {10.1086/117985}, \href
  {http://adsabs.harvard.edu/abs/1996AJ....112....9L} {112, 9}

\bibitem[\protect\citeauthoryear{{Lin} \& {Mohr}}{{Lin} \&
  {Mohr}}{2007}]{lin07}
{Lin} Y.-T.,  {Mohr} J.~J.,  2007, \mn@doi [\apjs] {10.1086/513565}, \href
  {http://adsabs.harvard.edu/abs/2007ApJS..170...71L} {170, 71}

\bibitem[\protect\citeauthoryear{{Lin}, {Mohr}  \& {Stanford}}{{Lin}
  et~al.}{2004}]{lin04a}
{Lin} Y.,  {Mohr} J.~J.,   {Stanford} S.~A.,  2004, \apj, \href
  {http://adsabs.harvard.edu/cgi-bin/nph-bib_query?bibcode=2004ApJ...610..745L&amp;db_key=AST}
  {610, 745}

\bibitem[\protect\citeauthoryear{{Lin}, {Partridge}, {Pober}, {Bouchefry},
  {Burke}, {Klein}, {Coish}  \& {Huffenberger}}{{Lin} et~al.}{2009}]{lin09}
{Lin} Y.,  {Partridge} B.,  {Pober} J.~C.,  {Bouchefry} K.~E.,  {Burke} S.,
  {Klein} J.~N.,  {Coish} J.~W.,   {Huffenberger} K.~M.,  2009, \mn@doi [\apj]
  {10.1088/0004-637X/694/2/992}, \href
  {http://adsabs.harvard.edu/abs/2009ApJ...694..992L} {694, 992}

\bibitem[\protect\citeauthoryear{{Lin}, {Shen}, {Strauss}, {Richards}  \&
  {Lunnan}}{{Lin} et~al.}{2010}]{lin10}
{Lin} Y.-T.,  {Shen} Y.,  {Strauss} M.~A.,  {Richards} G.~T.,   {Lunnan} R.,
  2010, \mn@doi [\apj] {10.1088/0004-637X/723/2/1119}, \href
  {http://adsabs.harvard.edu/abs/2010ApJ...723.1119L} {723, 1119}

\bibitem[\protect\citeauthoryear{{Lin}, {McDonald}, {Benson}  \&
  {Miller}}{{Lin} et~al.}{2015}]{linhenry15}
{Lin} H.~W.,  {McDonald} M.,  {Benson} B.,   {Miller} E.,  2015, \mn@doi [\apj]
  {10.1088/0004-637X/802/1/34}, \href
  {http://adsabs.harvard.edu/abs/2015ApJ...802...34L} {802, 34}

\bibitem[\protect\citeauthoryear{{Majumdar} \& {Mohr}}{{Majumdar} \&
  {Mohr}}{2003}]{majumdar03}
{Majumdar} S.,  {Mohr} J.~J.,  2003, \apj, \href
  {http://adsabs.harvard.edu/cgi-bin/nph-bib_query?bibcode=2003ApJ...585..603M&db_key=AST}
  {585, 603}

\bibitem[\protect\citeauthoryear{{Mantz}, {Allen}, {Ebeling}, {Rapetti}  \&
  {Drlica-Wagner}}{{Mantz} et~al.}{2010}]{mantz10b}
{Mantz} A.,  {Allen} S.~W.,  {Ebeling} H.,  {Rapetti} D.,   {Drlica-Wagner} A.,
   2010, \mn@doi [\mnras] {10.1111/j.1365-2966.2010.16993.x}, \href
  {http://adsabs.harvard.edu/abs/2010MNRAS.406.1773M} {406, 1773}

\bibitem[\protect\citeauthoryear{{Mauch}, {Murphy}, {Buttery}, {Curran},
  {Hunstead}, {Piestrzynski}, {Robertson}  \& {Sadler}}{{Mauch}
  et~al.}{2003}]{mauch03}
{Mauch} T.,  {Murphy} T.,  {Buttery} H.~J.,  {Curran} J.,  {Hunstead} R.~W.,
  {Piestrzynski} B.,  {Robertson} J.~G.,   {Sadler} E.~M.,  2003, \mnras, 342,
  1117

\bibitem[\protect\citeauthoryear{{Melin}, {Bartlett}  \&
  {Delabrouille}}{{Melin} et~al.}{2006}]{melin06}
{Melin} J.-B.,  {Bartlett} J.~G.,   {Delabrouille} J.,  2006, \mn@doi [\aap]
  {10.1051/0004-6361:20065034}, \href
  {http://adsabs.harvard.edu/abs/2006A%26A...459..341M} {459, 341}

\bibitem[\protect\citeauthoryear{{Merloni} et~al.,}{{Merloni}
  et~al.}{2012}]{merloni12}
{Merloni} A.,  et~al., 2012, preprint, \href
  {http://adsabs.harvard.edu/abs/2012arXiv1209.3114M} {} (\mn@eprint {arXiv}
  {1209.3114})

\bibitem[\protect\citeauthoryear{{Mills}}{{Mills}}{1981}]{mills81}
{Mills} B.~Y.,  1981, Proceedings of the Astronomical Society of Australia,
  \href {http://adsabs.harvard.edu/abs/1981PASAu...4..156M} {4, 156}

\bibitem[\protect\citeauthoryear{{Mocanu} et~al.,}{{Mocanu}
  et~al.}{2013}]{mocanu13}
{Mocanu} L.~M.,  et~al., 2013, \mn@doi [\apj] {10.1088/0004-637X/779/1/61},
  \href {http://adsabs.harvard.edu/abs/2013ApJ...779...61M} {779, 61}

\bibitem[\protect\citeauthoryear{{Mortonson}, {Hu}  \& {Huterer}}{{Mortonson}
  et~al.}{2011}]{mortonson11}
{Mortonson} M.~J.,  {Hu} W.,   {Huterer} D.,  2011, \mn@doi [\prd]
  {10.1103/PhysRevD.83.023015}, \href
  {http://adsabs.harvard.edu/abs/2011PhRvD..83b3015M} {83, 023015}

\bibitem[\protect\citeauthoryear{{Mullis} et~al.,}{{Mullis}
  et~al.}{2003}]{mullis03}
{Mullis} C.~R.,  et~al., 2003, \mn@doi [\apj] {10.1086/376866}, \href
  {http://adsabs.harvard.edu/abs/2003ApJ...594..154M} {594, 154}

\bibitem[\protect\citeauthoryear{{Murphy}, {Mauch}, {Green}, {Hunstead},
  {Piestrzynska}, {Kels}  \& {Sztajer}}{{Murphy} et~al.}{2007}]{murphy07}
{Murphy} T.,  {Mauch} T.,  {Green} A.,  {Hunstead} R.~W.,  {Piestrzynska} B.,
  {Kels} A.~P.,   {Sztajer} P.,  2007, \mn@doi [\mnras]
  {10.1111/j.1365-2966.2007.12379.x}, \href
  {http://adsabs.harvard.edu/abs/2007MNRAS.382..382M} {382, 382}

\bibitem[\protect\citeauthoryear{{Navarro}, {Frenk}  \& {White}}{{Navarro}
  et~al.}{1997}]{navarro97}
{Navarro} J.~F.,  {Frenk} C.~S.,   {White} S.~D.~M.,  1997, \mn@doi [\apj]
  {10.1086/304888}, \href {http://adsabs.harvard.edu/abs/1997ApJ...490..493N}
  {490, 493}

\bibitem[\protect\citeauthoryear{{Perlman}, {Horner}, {Jones}, {Scharf},
  {Ebeling}, {Wegner}  \& {Malkan}}{{Perlman} et~al.}{2002}]{perlman02}
{Perlman} E.~S.,  {Horner} D.~J.,  {Jones} L.~R.,  {Scharf} C.~A.,  {Ebeling}
  H.,  {Wegner} G.,   {Malkan} M.,  2002, \mn@doi [\apjs] {10.1086/339685},
  \href {http://adsabs.harvard.edu/abs/2002ApJS..140..265P} {140, 265}

\bibitem[\protect\citeauthoryear{{Piffaretti}, {Arnaud}, {Pratt},
  {Pointecouteau}  \& {Melin}}{{Piffaretti} et~al.}{2011}]{piffaretti11}
{Piffaretti} R.,  {Arnaud} M.,  {Pratt} G.~W.,  {Pointecouteau} E.,   {Melin}
  J.-B.,  2011, \mn@doi [\aap] {10.1051/0004-6361/201015377}, \href
  {http://adsabs.harvard.edu/abs/2011A%26A...534A.109P} {534, A109}

\bibitem[\protect\citeauthoryear{{Plagge} et~al.,}{{Plagge}
  et~al.}{2010}]{plagge10}
{Plagge} T.,  et~al., 2010, \mn@doi [\apj] {10.1088/0004-637X/716/2/1118},
  \href {http://adsabs.harvard.edu/abs/2010ApJ...716.1118P} {716, 1118}

\bibitem[\protect\citeauthoryear{{Planck Collaboration} et~al.,}{{Planck
  Collaboration} et~al.}{2011}]{planck11-13}
{Planck Collaboration} et~al., 2011, \mn@doi [\aap]
  {10.1051/0004-6361/201116471}, \href
  {http://adsabs.harvard.edu/abs/2011A%26A...536A..13P} {536, A13}

\bibitem[\protect\citeauthoryear{{Planck Collaboration} et~al.,}{{Planck
  Collaboration} et~al.}{2015b}]{planck15cosm}
{Planck Collaboration} et~al., 2015b, preprint, \href
  {http://adsabs.harvard.edu/abs/2015arXiv150201589P} {} (\mn@eprint {arXiv}
  {1502.01589})

\bibitem[\protect\citeauthoryear{{Planck Collaboration} et~al.,}{{Planck
  Collaboration} et~al.}{2015a}]{planck15}
{Planck Collaboration} et~al., 2015a, preprint, \href
  {http://adsabs.harvard.edu/abs/2015arXiv150201597P} {} (\mn@eprint {arXiv}
  {1502.01597})

\bibitem[\protect\citeauthoryear{{Pracy} et~al.,}{{Pracy}
  et~al.}{2016}]{pracy16}
{Pracy} M.,  et~al., 2016, preprint, \href
  {http://adsabs.harvard.edu/abs/2016arXiv160404332P} {} (\mn@eprint {arXiv}
  {1604.04332})

\bibitem[\protect\citeauthoryear{{Predehl} et~al.,}{{Predehl}
  et~al.}{2014}]{predehl14}
{Predehl} P.,  et~al., 2014, in Space Telescopes and Instrumentation 2014:
  Ultraviolet to Gamma Ray. p. 91441T, \mn@doi{10.1117/12.2055426}

\bibitem[\protect\citeauthoryear{{Reichardt} et~al.,}{{Reichardt}
  et~al.}{2013}]{reichardt13}
{Reichardt} C.~L.,  et~al., 2013, \mn@doi [\apj] {10.1088/0004-637X/763/2/127},
  \href {http://adsabs.harvard.edu/abs/2013ApJ...763..127R} {763, 127}

\bibitem[\protect\citeauthoryear{{Robertson}}{{Robertson}}{1991}]{robertson91}
{Robertson} J.~G.,  1991, Australian Journal of Physics, \href
  {http://adsabs.harvard.edu/abs/1991AuJPh..44..729R} {44, 729}

\bibitem[\protect\citeauthoryear{{Romer}, {Viana}, {Liddle}  \& {Mann}}{{Romer}
  et~al.}{2000}]{romer00}
{Romer} K.,  {Viana} P. T.~P.,  {Liddle} A.~R.,   {Mann} R.~G.,  2000, in Large
  Scale Structure in the X-ray Universe, Proceedings of the 20-22 September
  1999 Workshop, Santorini, Greece, eds. Plionis, M. \& Georgantopoulos, I.,
  Atlantisciences, Paris, France, p.409. p.~409, \url
  {http://adsabs.harvard.edu/cgi-bin/nph-bib_query?bibcode=2000lssx.proc..409R&db_key=AST}

\bibitem[\protect\citeauthoryear{{Saro} et~al.,}{{Saro} et~al.}{2014}]{saro14}
{Saro} A.,  et~al., 2014, \mn@doi [\mnras] {10.1093/mnras/stu575}, \href
  {http://adsabs.harvard.edu/abs/2014MNRAS.440.2610S} {440, 2610}

\bibitem[\protect\citeauthoryear{{Schaffer} et~al.,}{{Schaffer}
  et~al.}{2011}]{schaffer11}
{Schaffer} K.~K.,  et~al., 2011, \mn@doi [\apj] {10.1088/0004-637X/743/1/90},
  \href {http://adsabs.harvard.edu/abs/2011ApJ...743...90S} {743, 90}

\bibitem[\protect\citeauthoryear{{Schechter}}{{Schechter}}{1976}]{schechter76}
{Schechter} P.,  1976, \mn@doi [\apj] {10.1086/154079}, \href
  {http://adsabs.harvard.edu/abs/1976ApJ...203..297S} {203, 297}

\bibitem[\protect\citeauthoryear{{Sehgal}, {Bode}, {Das},
  {Hernandez-Monteagudo}, {Huffenberger}, {Lin}, {Ostriker}  \&
  {Trac}}{{Sehgal} et~al.}{2010}]{sehgal10}
{Sehgal} N.,  {Bode} P.,  {Das} S.,  {Hernandez-Monteagudo} C.,  {Huffenberger}
  K.,  {Lin} Y.,  {Ostriker} J.~P.,   {Trac} H.,  2010, \mn@doi [\apj]
  {10.1088/0004-637X/709/2/920}, \href
  {http://adsabs.harvard.edu/abs/2010ApJ...709..920S} {709, 920}

\bibitem[\protect\citeauthoryear{{Sehgal} et~al.,}{{Sehgal}
  et~al.}{2011}]{sehgal11}
{Sehgal} N.,  et~al., 2011, \mn@doi [\apj] {10.1088/0004-637X/732/1/44}, \href
  {http://adsabs.harvard.edu/abs/2011ApJ...732...44S} {732, 44}

\bibitem[\protect\citeauthoryear{{Staniszewski} et~al.,}{{Staniszewski}
  et~al.}{2009}]{staniszewski09}
{Staniszewski} Z.,  et~al., 2009, \mn@doi [\apj] {10.1088/0004-637X/701/1/32},
  \href {http://adsabs.harvard.edu/abs/2009ApJ...701...32S} {701, 32}

\bibitem[\protect\citeauthoryear{{Stocke}, {Perlman}, {Gioia}  \&
  {Harvanek}}{{Stocke} et~al.}{1999}]{stocke99}
{Stocke} J.~T.,  {Perlman} E.~S.,  {Gioia} I.~M.,   {Harvanek} M.,  1999,
  \mn@doi [\aj] {10.1086/300826}, \href
  {http://adsabs.harvard.edu/abs/1999AJ....117.1967S} {117, 1967}

\bibitem[\protect\citeauthoryear{{Sunyaev} \& {Zel'dovich}}{{Sunyaev} \&
  {Zel'dovich}}{1972}]{sunyaev72}
{Sunyaev} R.~A.,  {Zel'dovich} Y.~B.,  1972, Comments on Astrophysics and Space
  Physics, \href
  {http://adsabs.harvard.edu/cgi-bin/nph-bib_query?bibcode=1972CoASP...4..173S&amp;db_key=AST}
  {4, 173}

\bibitem[\protect\citeauthoryear{{Tegmark} \& {de Oliveira-Costa}}{{Tegmark} \&
  {de Oliveira-Costa}}{1998}]{tegmark98}
{Tegmark} M.,  {de Oliveira-Costa} A.,  1998, \mn@doi [\apjl] {10.1086/311410},
  \href {http://adsabs.harvard.edu/abs/1998ApJ...500L..83T} {500, L83}

\bibitem[\protect\citeauthoryear{{Tinker}, {Kravtsov}, {Klypin}, {Abazajian},
  {Warren}, {Yepes}, {Gottl{\"o}ber}  \& {Holz}}{{Tinker}
  et~al.}{2008}]{tinker08}
{Tinker} J.,  {Kravtsov} A.~V.,  {Klypin} A.,  {Abazajian} K.,  {Warren} M.,
  {Yepes} G.,  {Gottl{\"o}ber} S.,   {Holz} D.~E.,  2008, \mn@doi [\apj]
  {10.1086/591439}, \href {http://adsabs.harvard.edu/abs/2008ApJ...688..709T}
  {688, 709}

\bibitem[\protect\citeauthoryear{{Vanderlinde} et~al.,}{{Vanderlinde}
  et~al.}{2010}]{vanderlinde10}
{Vanderlinde} K.,  et~al., 2010, \mn@doi [\apj] {10.1088/0004-637X/722/2/1180},
  \href {http://adsabs.harvard.edu/abs/2010ApJ...722.1180V} {722, 1180}

\bibitem[\protect\citeauthoryear{{Vieira} et~al.,}{{Vieira}
  et~al.}{2010}]{vieira10}
{Vieira} J.~D.,  et~al., 2010, \mn@doi [\apj] {10.1088/0004-637X/719/1/763},
  \href {http://adsabs.harvard.edu/abs/2010ApJ...719..763V} {719, 763}

\bibitem[\protect\citeauthoryear{{Vikhlinin} et~al.,}{{Vikhlinin}
  et~al.}{2009}]{vikhlinin09b}
{Vikhlinin} A.,  et~al., 2009, \mn@doi [\apj] {10.1088/0004-637X/692/2/1033},
  \href {http://adsabs.harvard.edu/abs/2009ApJ...692.1033V} {692, 1033}

\bibitem[\protect\citeauthoryear{{de Haan} et~al.,}{{de Haan}
  et~al.}{2016}]{dehaan16}
{de Haan} T.,  et~al., 2016, preprint, \href
  {http://adsabs.harvard.edu/abs/2016arXiv160306522D} {} (\mn@eprint {arXiv}
  {1603.06522})

\bibitem[\protect\citeauthoryear{{de Zotti}, {Ricci}, {Mesa}, {Silva},
  {Mazzotta}, {Toffolatti}  \& {Gonz{\'a}lez-Nuevo}}{{de Zotti}
  et~al.}{2005}]{dezotti05}
{de Zotti} G.,  {Ricci} R.,  {Mesa} D.,  {Silva} L.,  {Mazzotta} P.,
  {Toffolatti} L.,   {Gonz{\'a}lez-Nuevo} J.,  2005, \mn@doi [\aap]
  {10.1051/0004-6361:20042108}, \href
  {http://adsabs.harvard.edu/abs/2005A%26A...431..893D} {431, 893}

\makeatother
\end{thebibliography}

\end{document}